\documentclass{ieeeaccess}

\usepackage[utf8]{inputenc}
\usepackage[T1]{fontenc}
\usepackage{braket}
\usepackage{amsthm}

\newcommand{\eqdef}{\stackrel{\triangle}{=}}

\newcommand{\ketbra}[2]{\vert #1 \rangle  \langle #2 \vert}

\usepackage{mathtools, cuted}
\usepackage{mathtools}
\usepackage{subcaption}
\usepackage{cite}
\usepackage{amssymb,amsfonts}
\usepackage{algorithmic}
\usepackage{graphicx}
\usepackage{textcomp}
\usepackage{xcolor}
\usepackage{url}
\usepackage{hyperref}

\newtheorem{theorem}{\textbf{Theorem}}
\newtheorem{proposition}{\textbf{Proposition}}

\theoremstyle{remark}
\newtheorem*{remark}{\textbf{Remark}}

\usepackage{cite}
\usepackage{algorithmic}
\usepackage{graphicx}
\usepackage{textcomp}
\def\BibTeX{{\rm B\kern-.05em{\sc i\kern-.025em b}\kern-.08em
    T\kern-.1667em\lower.7ex\hbox{E}\kern-.125emX}}
\begin{document}
\history{}
\doi{}

\title{Deterministic Generation of Multipartite Entanglement via Causal Activation \\in the Quantum Internet}
\author{\uppercase{Seid Koudia}\authorrefmark{1,2}\authorrefmark{*},
\uppercase{Angela Sara Cacciapuoti}\authorrefmark{1,3}\authorrefmark{†}, \IEEEmembership{Senior Member, IEEE}, and Marcello Caleffi\authorrefmark{1,3}\authorrefmark{‡},
\IEEEmembership{Senior Member, IEEE}}

\address[1]{\textit{Quantum Internet} research group @ \textit{FLY: Future Communications Laboratory}, Department of Electrical Engineering and Information Technology (DIETI), University of Naples \textit{Federico II}, Naples, 80125 Italy. \href{http://www.quantuminternet.it}{www.quantuminternet.it}}
\address[2]{Department of Physics ``Ettore Pancini'', University of Naples \textit{Federico II}, Naples, 80125 Italy. }
\address[3]{Laboratorio Nazionale di Comunicazioni Multimediali, \textit{CNIT: National Inter-University Consortium for Telecommunications}, Naples, 80126 Italy}
\address[*]{e-mail: seid.koudia@unina.it}
\address[†]{e-mail: e-mail: angelasara.cacciapuoti@unina.it}
\address[‡]{e-mail: e-mail: marcello.caleffi@unina.it}

\corresp{Corresponding author: Angela Sara Cacciapuoti (e-mail: angelasara.cacciapuoti@unina.it).}

\begin{abstract}
Entanglement represents ``\textit{the}'' key resource for several applications of quantum information processing, ranging from quantum communications to distributed quantum computing. Despite its fundamental importance, deterministic generation of maximally entangled qubits represents an on-going open problem. Here, we design a novel generation scheme exhibiting two attractive features, namely, i) deterministically generating \textcolor{black}{different classes -- particularly, GHZ-like, W-like and graph states -- of} genuinely multipartite entangled states, ii) not requiring any direct interaction between the qubits. Indeed, the only necessary condition is the possibility of coherently controlling -- 
according to the indefinite causal order framework -- the causal order among \textcolor{black}{the} unitaries acting on the qubits. Through the paper, we analyze and derive the conditions on the unitaries for deterministic generation, and we provide examples for unitaries practical implementation. We conclude the paper by discussing the scalability of the proposed scheme to higher dimensional \textcolor{black}{genuine multipartite entanglement (GME)} states and by introducing some possible applications of the proposal for quantum networks.
\end{abstract}


\titlepgskip=-21pt
\maketitle

\section{Introduction}
\label{Sec:01}
One of the most fundamental concepts within the quantum realm is the notion of quantum entanglement. It is well established that entangled states -- even in the simplest form of two-qubit entangled states -- are essential to enable the marvels of quantum information processing \cite{CacCalTafCatGherBia2020,IllCalMan-22,Pirandola2016UniteTB,Wehner2018QuantumIA,Wilde,nielsen_chuang_2010} within the Quantum Internet. And, as a matter of fact, both the theory of entanglement and its experimental generation have been a topic of intensive research.\\
Indeed, several applications of quantum information processing -- ranging from quantum communications through distributed quantum key distribution to distributed quantum computing -- rely on the generation and the remote distribution of entangled flying qubits \cite{Flhmann2019EncodingAQ,arute2019quantum,boixo2018characterizing,lund2017quantum}, with a wide consensus within the research community on light being the ideal substrate for quantum information carriers. Nevertheless, given the limitations of current schemes for photonic entanglement generation, the research is still ongoing. In fact, some of the available schemes are probabilistic \cite{WanCheLuo_2016,RosAraFabTre_2013,CaiRosFerArz_2017}, relying as instance on some form of parametric down conversion. Other schemes require a tight matter-flying interaction \cite{Besse_2020,Schwartz_2016}. Clearly, when it comes to multi-partite entanglement, both the approaches hardly scale to large systems. This has driven a recent interest in designing all-photonic deterministic sources of entanglement \cite{Istrati_2020,Takeda_2019}.\\
In this work, we contribute toward this research direction by resorting to a recently proposed framework for quantum information processing, namely, the superposition of causal orders \cite{3,8,10,13,18,19,20,37,38,83,Rubino_2017,Procopio_2015,procopio-2020,40,koudia-2021,koudia2021quantum,koudia2019superposition,Miao,white,WeiJian, guha2022quantum}. Specifically, we design an entanglement generation scheme where a superposition of causal orders between local unitaries, acting on qubits in pure product states, deterministically generates genuinely multipartite entangled (GME) states. Interestingly, the proposed scheme efficiently scales to higher dimensional GME states, due to the simplicity and the modularity of the protocol architecture. Furthermore, the scheme does not require any direct interaction among the input qubits or between the input qubit and the qubit governing the quantum control of the causal order between the unitaries. Indeed, the only requirement is the possibility of coherently controlling the causal order among the unitaries.\\
It is worthwhile to note that -- by exploiting the super-map formalism -- the design of the proposed scheme has been conducted without any specific assumption on the particulars of the underlying qubit technology. However, when it comes to practical implementation, we can recognize that   
the proposed scheme for deterministic entanglement generation is achievable in near-term quantum networks, as coherent control of causal orders is affordable by current technology level and it has been successfully implemented for flying qubits \cite{Rubino_2017,Procopio_2015,white}. \textcolor{black}{From a resource theoretic point of view \cite{chiribella_2020,Milz-2021}, we are assuming that single qubit unitaries are given as a free resource. Furthermore, we are assuming coherent control of unitaries -- with CNOT representing the pivotal example -- not as a free resource, but rather as unavailable since it may be very difficult to implement with the available quantum technology, as it happens with photonic platforms.}
In this light, through the paper, we discuss some possible applications of the proposal to quantum networks. Specifically, we recognize that the coherent control of the causal order of collective single qubit unitaries -- either locally on particular nodes or globally on some given cluster of nodes -- constitutes a novel paradigm to generate and distribute multipartite entangled states among remote nodes and in a scalable way. \textcolor{black}{Importantly, we will show that our scheme can be used to establish  resourceful states for measurement based quantum computation\cite{Hein-2006} remotely within the quantum network.}

\section{Background and Notation}
\label{Sec:2.1}

\begin{figure*}[t]
\color{black}
    \begin{minipage}[c] {0.49\textwidth}
        \begin{center}
            \includegraphics[width=1\columnwidth ]{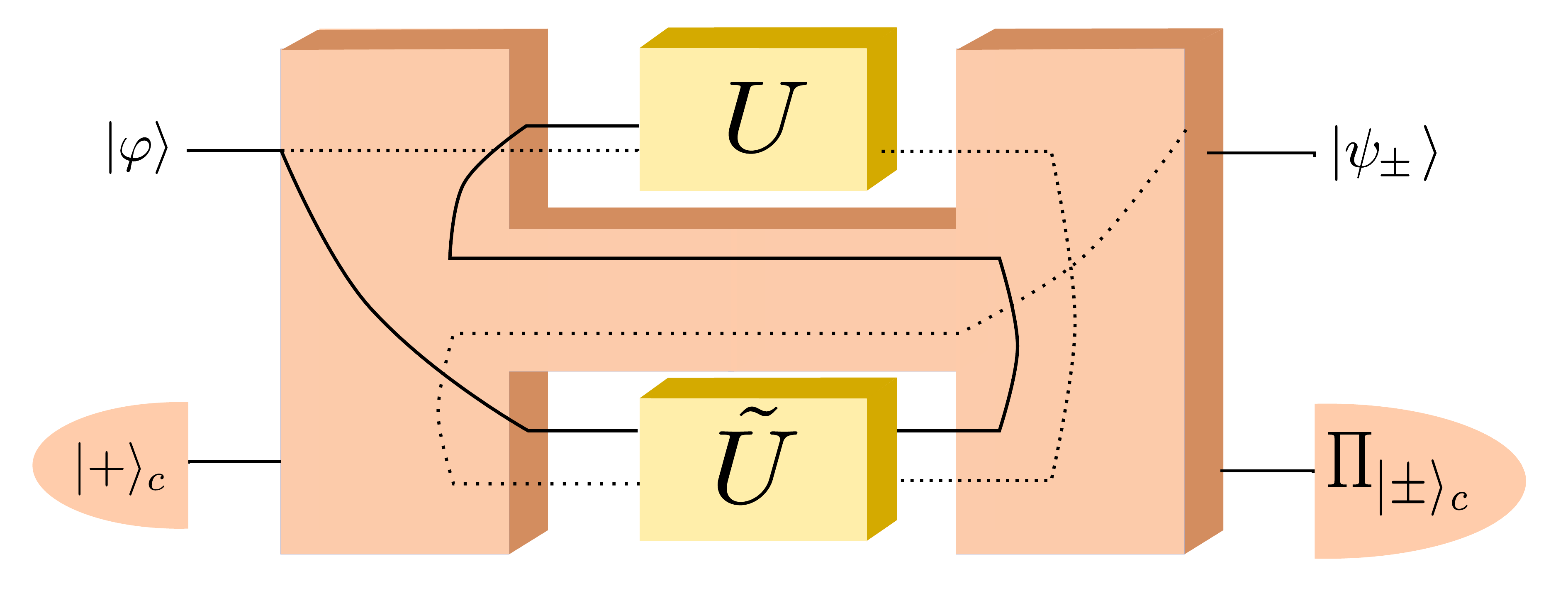}
        \end{center}
        \subcaption{\color{black}Graphical representation of the quantum switch placing the unitaries $U$ and $\tilde{U}$ in an even superposition of causal orders. The control qubit, viewed as a fixed parameter of the communication network between the sender and the receiver and set to $\ket{+}$, becomes part of the supermap responsible for the coherent control of the causal orders and it is shown in the bottom part of the graph. By measuring the control qubit in the coherent basis  $\{\ket{-},\ket{+}\}$, one of the two alternative states $\ket{\psi_{\pm}}$ emerges as output.}
        \label{Fig:01-a}
    \end{minipage}
    \begin{minipage}[c] {0.49\textwidth}
        \begin{center}
            \includegraphics[width=1\columnwidth]{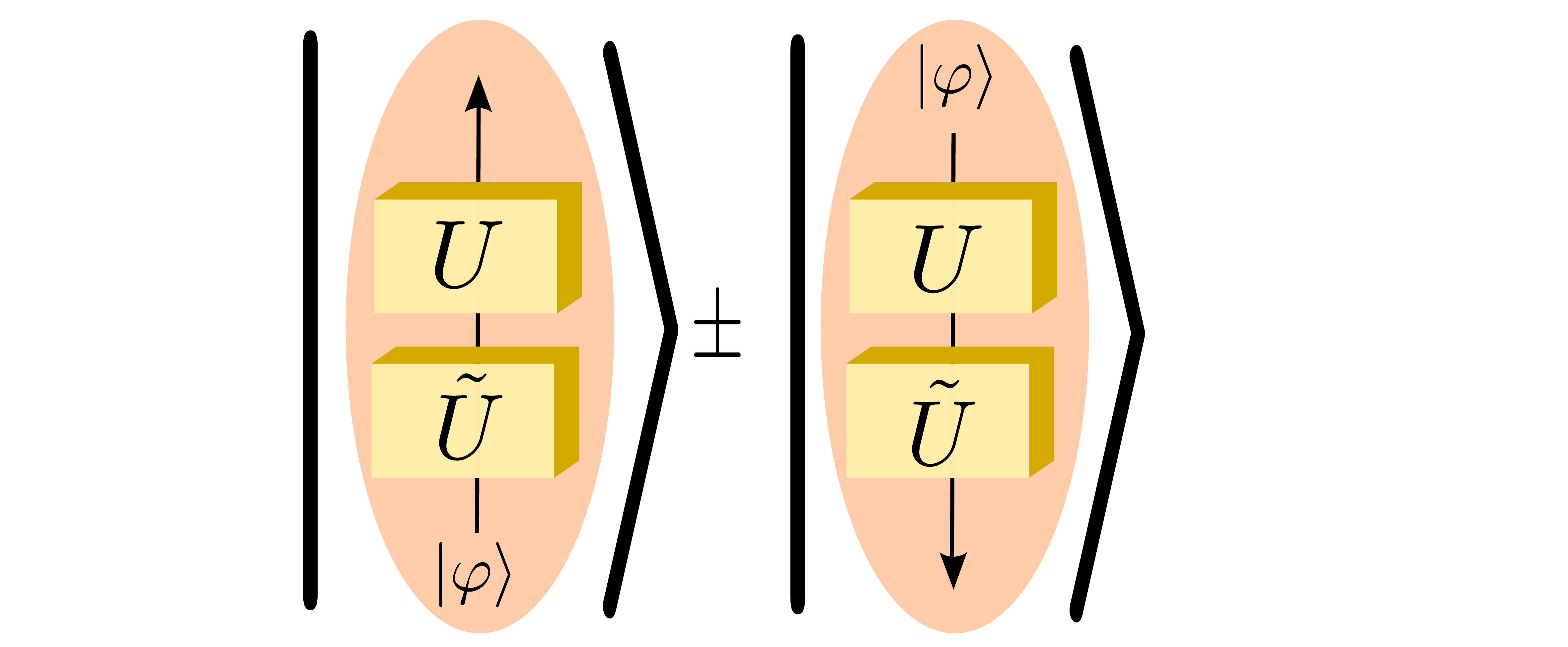}
        \end{center}
        \subcaption{\color{black}Pictorial representation of the output state $\ket{\psi}_{\pm}$, once the control qubit is measured. The output $\ket{\psi_{\pm}}$ is given by the coherent superposition of two terms. Within the first one, the unitaries operate on the input according to the causal order where $\tilde{U}$ is applied before $U$, whereas within the second one the unitaries operates following the alternative order where $U$ is applied before $\tilde{U}$.}
        \label{Fig:01-b}
    \end{minipage}
    \caption{\color{black}Quantum Switch implementing an even superposition of the two alternative causal orders between two unitaries $U$ and $\tilde{U}$ operating on the input state $\ket{\varphi}$.}
    \label{Fig:01}
    \hrulefill
\end{figure*}

With a series of recent works, researchers have shown that quantum placement of quantum channels -- namely, placing quantum channels in a coherent superposition of alternative configurations -- can provide significant advantages for a number of problems, ranging from quantum computation \cite{ColDarFac-12,83,AraCosBru-14} and quantum information processing \cite{Chi-12,WakSoeMur-19} through non-local games \cite{OreCosBru-12} to communication complexity \cite{FeiAraBru-15,GueFeiAra-16,cacciapuoti2019capacity,marcello,CalCacBia2018,daryus,daryus0}. Instances of this quantum placement range from superposition of alternative quantum channels, traversed by the information carrier, to superposition of alternative causal orders between the quantum channels. With reference to the superposition of causal orders between quantum channels, the placement is realized through an higher-order map known as \textit{quantum switch} \cite{83}.\\
\textcolor{black}{Mathematically, the quantum switch is described by a supermap $S$ taking two channels $U(\cdot)$ and $\tilde{U}(\cdot)$ as inputs, and giving as output a channel resulting from the combination of $U(\cdot)$ and $\tilde{U}(\cdot)$ in a superposition of causal orders, controlled by a quantum degree of freedom $\ket{\varphi_c}$. Its action on quantum states is defined by the Kraus operators\cite{83,3} $S_{ij}= U_i \tilde{U}_j \otimes \ketbra{0}{0}_c + \tilde{U}_j U_i \otimes \ketbra{1}{1}_c$, where $\{ U_i \}$ and $\{\tilde{U}_j\}$ denote the Kraus operators of the primitive channels $U(\cdot)$ and $\tilde{U}(\cdot)$, and $\{\ket{0}_c, \ket{1}_c\}$ denotes the orthogonal states of the control system. Accordingly, the resultant channel implemented by the quantum switch is given by:
\begin{equation}
    \label{Eq:01}
    S(U,\tilde{U})(\rho\otimes \rho_c) = \sum_{ij} S_{ij}(\rho \otimes \rho_c) S_{ij}^{\dagger},
\end{equation}
with $\rho$ and $\rho_c$ denoting the density matrix of the input and the control, respectively. Indeed, many channels, beyond the two channels scenario, can be fed into the quantum switch supermap as we will see in the next sections \cite{3, chiribella_2020}.}\\
\textcolor{black}{In the following, giving that we aim at generating maximally entangled states, we focus on pure input states and unitary channels, therefore it is needless to go through the density matrix formalism of Eq.~\ref{Eq:01}}. Furthermore, we set the control $\ket{\varphi_c}$ to $\ket{+}$, i.e., we place the primitive channels in an even superposition of causal orders, accordingly to \cite{cacciapuoti2019capacity, white,83,10,13,18}. Accordingly, the quantum switch supermap in \eqref{Eq:01} exhibits a single unitary Kraus operator $S = U \tilde{U} \otimes \ketbra{0}{0}_c + \tilde{U} U \otimes \ketbra{1}{1}_c$, with $U$ and $\tilde{U}$ denoting the (single) Kraus operators of the primitive channels, leading to the overall operation on the input state $\ket{\varphi}$ given by\footnote{The notation has been simplified with respect to the one in \eqref{Eq:01} to highlight the focus on pure input states and unitary channels.}:
\begin{align}
    \label{Eq:02}
     S(\ket{\varphi}\otimes \ket{\varphi_c})=&
     \frac{1}{2}\big( U \tilde{U} + \tilde{U} U \big) \ket{\varphi} \otimes \ket{+}_c +\nonumber\\
     &+\frac{1}{2}\big( U \tilde{U} - \tilde{U} U \big) \ket{\varphi} \otimes \ket{-}_c
\end{align}
After performing a measurement on the control qubit in the coherent basis, the following outcome states -- highlighting the superposition of causal orders between the unitaries -- emerge:
\begin{equation}
    \label{Eq:03}
    \ket{\psi_\pm} =
        \frac{1}{\sqrt{L}_\pm} \big( U \tilde{U} \pm \tilde{U} U \big) \ket{\varphi} =
        \frac{1}{\sqrt{L}_\pm} \big( \overrightarrow{U} \pm \overleftarrow{U} \big) \ket{\varphi}
\end{equation}
where $L_\pm$ is a normalization constant, depending on both the unitaries $U, \tilde{U}$ and on the postselected state. In \eqref{Eq:03}, we introduced $\overrightarrow{U},\overleftarrow{U}$ as a shorthand notation, with $\overrightarrow{U} \eqdef U \tilde{U}$ denoting the order where $\tilde{U}$ is applied before $U$ and $\overleftarrow{U} \eqdef \tilde{U} U$ denoting the alternative order. This is schematized in Fig.~\ref{Fig:01} where we omitted -- as extensively done whenever possible through the rest of the paper -- the normalization constant for the sake of simplicity. Indeed, from Fig.~\ref{Fig:01-b}, it is intuitive to grasp that, once the control qubit is measured, the output is a coherent superposition of two \textcolor{black}{contributions}, where in each \textcolor{black}{contribution} the unitaries process the input according to one of the two alternative causal orders, namely,  either $\overleftarrow{U}$ or $\overrightarrow{U}$.

\textcolor{black}{It is worth mentioning that Eq.~\eqref{Eq:03} can be easily simulated by a quantum circuit with fixed causal order, as long as a coherent control among unitaries acting on different qubits -- as instance, a CNOT gate -- is available \cite{39,Daniele-2021}. Nevertheless, controlled gates are not easy to realize in photonic platforms, in contrast to superconducting technologies among other ones. Accordingly, the aim of our manuscript is to rather do the opposite of assuming the availability of CNOT gates. In fact, it aims at investigating whether indefinite causal order of local unitaries  can generate some sort of entanglement.}

\section{Results}
\label{Sec:02}

\subsection*{Bell states generation}
\label{Sec:2.2}

\begin{figure*}[t]
    \begin{minipage}{0.48\textwidth}
        \begin{center}
            \includegraphics[width=1\columnwidth]{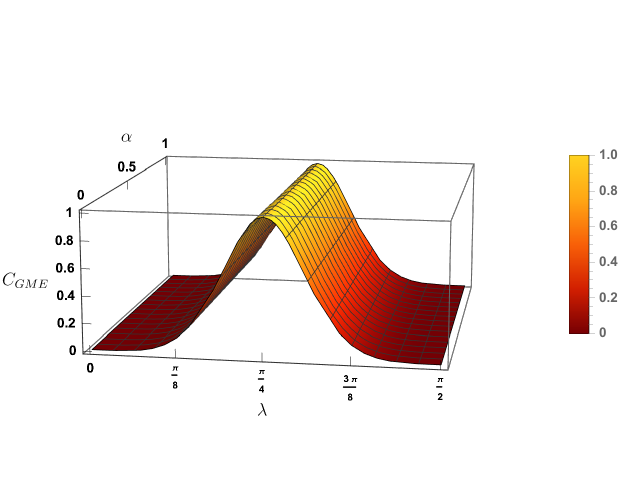}
        \end{center}
        \subcaption{Concurrence $C(\rho_+)$ of the state $\ket{\psi^{(2)}_+}$ given in \eqref{Eq:04}, obtained as output when the control qubit is measured as $\ket{+}$ with $\rho_+ \eqdef \ket{\psi^{(2)}_+}\bra{\psi^{(2)}_+}$.}
        \label{Fig:03-a}
    \end{minipage}
    \hspace{0.03\textwidth}
    \begin{minipage}{0.48\textwidth}
        \begin{center}
            \includegraphics[width=1\columnwidth]{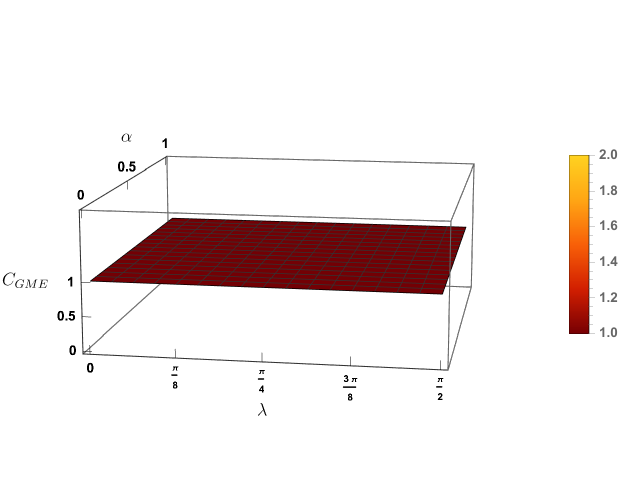}
        \end{center}
        \subcaption{Concurrence $C(\rho_-)$ of the state $\ket{\psi^{(2)}_-}$ given in \eqref{Eq:04}, obtained as output when the control qubit is measured as $\ket{-}$ with $\rho_- \eqdef \ket{\psi^{(2)}_-}\bra{\psi^{(2)}_-}$.}
        \label{Fig:03-b}
    \end{minipage}
    \caption{3D plot for the concurrence of the output states $\ket{\psi^{(2)}_{\pm}}$ given in \eqref{Eq:04} as function of: i) the y-rotation parameter $\lambda$ controlling the unitaries $\tilde{U_0}$ and $\tilde{U_1}$, and ii) the superposition parameter $\alpha$ controlling the input state $\ket{\varphi_0 \varphi_1}$, assumed real. Unitaries $U_0$ and $U_1$ both set to a Pauli-z gate. Maximally entangled states exhibit unitary concurrence.}
    \label{Fig:03}
    \hrulefill
\end{figure*}

Let us now consider two local unitary operators $V^{(2)} = U_0 \otimes U_1$ and $\tilde{V}^{(2)} = \tilde{U}_0 \otimes \tilde{U}_1$ and a 2-qubit input system in the separable state $\ket{\varphi_0} \otimes \ket{\varphi_1}$.\\
Being the input in a product state and given the assumption of local unitaries, the resulting outcome, for any causal order between the local unitaries such as $V^{(2)}\tilde{V}^{(2)}$ or $\tilde{V}^{(2)}V^{(2)}$, will be a product state as well. Furthermore, no entanglement can be distilled from such a state in the asymptotic limit with two-way \textcolor{black}{local operations and classical communication (LOCC)} assistance \cite{Horodecki_2009}.\\
Conversely, if we process the separable input through an even superposition of causal orders between the two unitaries \textcolor{black}{-- similarly to the scheme shown in Fig.~\ref{Fig:01-a} --} by measuring the control in the coherent basis, the following output
emerges:
\begin{equation}
    \label{Eq:04}
    \ket{\psi^{(2)}_{\pm}} = \frac{1}{\sqrt{L}_\pm} 
        \big(V^{(2)} \tilde{V}^{(2)} \pm \tilde{V}^{(2)} V^{(2)} \big) \ket{\varphi_0 \varphi_1}
\end{equation}
From \eqref{Eq:04} we note that, once the control qubit is measured, the output is a coherent superposition of two \textcolor{black}{contributions}, where in each \textcolor{black}{contribution} the unitaries process the separable input according to one of the two alternative -- i.e., either $V^{(2)} \tilde{V}^{(2)}$ or $\tilde{V}^{(2)} V^{(2)}$ -- causal orders. \textcolor{black}{This can be schematized in a similar way to Figure~\ref{Fig:01-b} with a bipartite separable initial state and bipartite local unitaries}.\\
Now, the main question arises: is there any entanglement within the quantum state $\ket{\psi^{(2)}_{\pm}}$ emerging out of the controlled superposition of causal orders? The answer to this question is definitely yes. Indeed, the output state is maximally entangled if and only if the following condition on the local unitaries $\{U_i\,, \tilde{U}_i\}_{i=0,1}$ holds (see Theorem~\ref{Theo:01} in Appendix.~\ref{appendix.2}): 
\begin{equation}
    \label{Eq:05}
	\bra{\varphi_0} \overleftarrow{U_0}^{\dagger} \overrightarrow{U_0} \ket{\varphi_0} = 0
	\quad \wedge \quad
	\bra{\varphi_1} \overleftarrow{U_1}^{\dagger} \overrightarrow{U_1} \ket{\varphi_1}=0
\end{equation}
with $\wedge$ denoting the Boolean operator $\texttt{AND}$, and with $\overleftarrow{\cdot},\overrightarrow{\cdot}$ being the introduced shorthand notations for the alternative causal orders among the unitaries, i.e., $\overrightarrow{U_i} \eqdef U_i \tilde{U_i}$ and $\overleftarrow{U_i} \eqdef \tilde{U_i} U_i$ for $i \in \{0,1\}$. \\
Stemming from this result, we derive a \textit{lighter} condition assuring the separability of the output state (see Proposition~\ref{Prop:01} in Appendix.~\ref{appendix.2}). Specifically, if there exists at least one $i \in \{0,1\}$ so that $\bra{\varphi_i} \overleftarrow{U_i}^{\dagger} \overrightarrow{U_i} \ket{\varphi_i} = 1$, the output state in \eqref{Eq:04} is separable.\\
Clearly, one could wonder which are the requirements in terms of unitaries and input state so that condition \eqref{Eq:05} can be satisfied. Namely, how ``easily'' entanglement can be obtained out of a superposition of causal orders.\\
To address this crucial aspect, we note that the sufficient and necessary condition in \eqref{Eq:05} consists of two \textit{separate} constraints, with the former constraint operating only on $U_0, \tilde{U_0}$ and $\ket{\varphi_0}$ whereas the latter one depends only on $U_1, \tilde{U_1}$ and $\ket{\varphi_1}$. This separability feature allows us to design unitaries $U_0$ and $\tilde{U_0}$ independently from $U_1, \tilde{U_1}$. Furthermore, the condition in \eqref{Eq:05} can be satisfied with practical unitaries, as shown in the following.\\
To this aim, let us assume as input state $\ket{\varphi_0 \varphi_1} = \ket{\eta \eta}$, with $\ket{\eta} = \sqrt{\alpha} \ket{0} + \sqrt{1-\alpha} \ket{1}$ being an arbitrary superposition of basis states. Furthermore, let us assume both $U_0$ and $U_1$ representing the popular Pauli-z gate, i.e., $U_i = \sigma_z$. Finally, let us assume both $\tilde{U_0}$ and $\tilde{U_1}$ being the y-rotation gate $R_y(2\lambda) = e^{-i \sigma_y \lambda}$, with $\sigma_y$ denoting the Pauli-y gate. Let's now consider the two possible events, namely, control qubit measured either as $\ket{+}$ or as $\ket{-}$. In the former case, the condition for entangled output state given in \eqref{Eq:04}
translates to $\lambda \neq 0, \frac{\pi}{2}$. In fact, only when the unitary parameter $\lambda$ is either equal to $0$ or $\frac{\pi}{2}$, \textcolor{black}{the output state generated by the quantum switch is separable}. Furthermore, the condition for maximally entangled output given in \eqref{Eq:05} is translated to $\lambda = \frac{\pi}{4}$. \textcolor{black}{This is shown in Figure~\ref{Fig:03-a} by plotting the concurrence (see Appendix.~\ref{appendix.1}), which is an entanglement measure that can fully characterize the entanglement content of the output state $\ket{\psi^{(2)}_+}$ as a function of the parameters, i.e., the concurrence is maximum and equals one when the state is maximally entangled, and it decreases monotonically with the decrease of the entanglement content of the state until reaching its vanishing point.}. As regards to the latter case -- namely, whenever the control qubit is measured as $\ket{-}$ -- the output is either separable or maximally entangled. And the state is separable only if $\lambda$ is either equal to $0$ or $\frac{\pi}{2}$, whereas it is maximally entangled for any $\lambda$ in $(0,\frac{\pi}{2})$, as shown in Figure~\ref{Fig:03-b}. \textcolor{black}{It is worth-noting that another important feature that the concurrence in Figure~\ref{Fig:03-b} is highlighting, is the fact that even if the single qubit unitaries fails to meet the optimality requirement for deterministic Bell pairs generation, i.e., $\lambda=\frac{\pi}{4}$, probabilistic Bell pair generation is always possible within the range $0<\lambda<\frac{\pi}{2}$.}  \\
We have shown that the scheme can be implemented through a coherent control of straightforward unitaries: the Pauli-z gate and the y-rotation gate $R_y(\frac{\pi}{2})$.  \textcolor{black}{It is important to note that an equivalent example has been given independently \cite{zych-2019,Rubino-2022}, by studying different foundational contexts of the indefinite causal order framework.}

\subsection*{GHZ-like states generation}
\label{Sec:2.3}

\begin{figure*}[t]
    \begin{minipage}[c] {0.48\textwidth}
        \begin{center}
            \includegraphics[width=1\columnwidth]{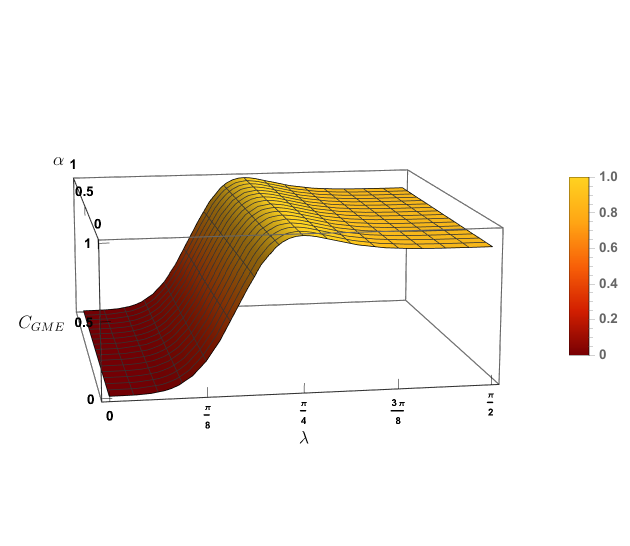}
        \end{center}
        \subcaption{GME concurrence $C_{\text{GME}}(\rho_+)$ of the state $\ket{\psi^{(3)}_+}$ given in \eqref{Eq:06}, obtained as output when the control qubit is measured as $\ket{+}$ with $\rho_+ \eqdef \ket{\psi^{(3)}_+}\bra{\psi^{(3)}_+}$.}
        \label{Fig:05-a}
    \end{minipage}
    \hspace{0.03\textwidth}
    \begin{minipage}{0.48\textwidth}
        \begin{center}
            \includegraphics[width=1\columnwidth]{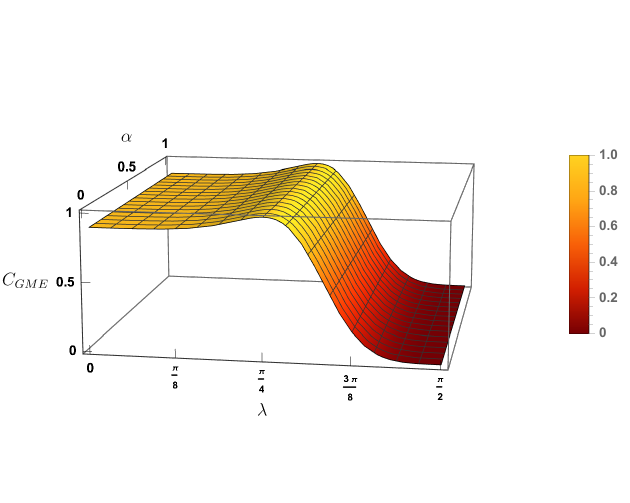}
        \end{center}
        \subcaption{GME concurrence $C_{\text{GME}}(\rho_-)$ of the state $\ket{\psi^{(3)}_-}$ given in \eqref{Eq:06}, obtained as output when the control qubit is measured as $\ket{-}$ with $\rho_- \eqdef \ket{\psi^{(3)}_-}\bra{\psi^{(3)}_-}$.}
        \label{Fig:05-b}
    \end{minipage}
    \caption{3D plot for the GME concurrence of the output state $\ket{\psi^{(3)}_{\pm}}$ given in \eqref{Eq:06} as function of: i) the y-rotation parameter $\lambda$ controlling the unitaries $\{ \tilde{U_i} \}$, and ii) the superposition parameter $\alpha$ controlling the input state $\ket{\varphi_0 \varphi_1 \varphi_2}$, assumed real. Unitaries $\{ U_i \}$ set to Pauli-z gate. GHZ-like states exhibit unitary GME concurrence.}
    \label{Fig:05}
    \hrulefill
\end{figure*}
Similarly to the bipartite case, we consider an even superposition of the two alternative causal orders between two 3-qubit local unitaries $V^{(3)} = U_0 \otimes U_1 \otimes U_2$ and $\tilde{V}^{(3)}= \tilde{U}_0 \otimes \tilde{U_1} \otimes \tilde{U_2}$ acting on an initially pure product tripartite state $\ket{\varphi_0 \varphi_1 \varphi_2}$, \textcolor{black}{similarly to the scheme shown in Figure~\ref{Fig:01-a}}. By measuring the control qubit in the coherent basis and according to eq.~\eqref{Eq:03}, we obtain the following output state:
\begin{equation}
    \ket{\psi^{(3)}_\pm} = \frac{1}{\sqrt{L}_\pm} 
        \big(V^{(3)} \tilde{V}^{(3)} \pm \tilde{V}^{(3)} V^{(3)} \big) \ket{\varphi_0 \varphi_1 \varphi_2}
    \label{Eq:06}
\end{equation}
Indeed, as in the bipartite case, the output is a superposition of two different input processing, with the two \textcolor{black}{processes} differing for the causal order between the unitaries. This similarity maps as well into the necessary and sufficient condition for the output in \eqref{Eq:06} being a GHZ-like state, which is given by (see Theorem~\ref{Theo:02} in Appendix.~\ref{appendix.3}):
\begin{equation}
\bra{\varphi_i} \overleftarrow{U_i}^{\dagger} \overrightarrow{U_i} \ket{\varphi_i} = 0 \quad \forall  i = 0,1,2
\label{Eq:07}
\end{equation}
and $\overleftarrow{\cdot},\overrightarrow{\cdot}$ being the usual shorthand notations for the alternative causal orders among the unitaries. Hence, the output in \eqref{Eq:06} is a legitimate GHZ-like state if and only if $\overleftarrow{U_i} \ket{\varphi_i}$ is orthonormal to $\overrightarrow{U_i} \ket{\varphi_i}$.\\
Indeed, there exists a lighter condition (see Proposition~\ref{Prop:02} in Appendix.~\ref{appendix.3} assuring the separability of the output state \eqref{Eq:06}, given by:
\begin{equation}
    \exists \, i \in \{0,1,2\} \; : \; \overleftarrow{U_i} \ket{\varphi_i} = \overrightarrow{U_i} \ket{\varphi_i}
	\label{Eq:08}
\end{equation}
\textcolor{black}{It is crucial to note that this straightforwardly extends to $n$-partite GHZ-like states by considering two $n$-qubit local unitaries $V^{(n)}$ and $\tilde{V}^{(n)}$ acting on a $n$-partite separable state.} In such a case, the output is a legitimate GHZ-like state as long as (see Remark following Theorem~\ref{Theo:02} in Appendix.~\ref{appendix.3}):
\begin{equation}
    \bra{\varphi_i} \overleftarrow{U_i}^{\dagger} \overrightarrow{U_i} \ket{\varphi_i} = 0 \quad \forall \, i = 0, \ldots, n-1
	\label{Eq:09}
\end{equation}
Clearly, the higher is the \textit{dimension} of the GHZ-like state to be generated, the higher is the number of constraints in \eqref{Eq:09} that must be simultaneously satisfied. However, this is not an issue, given that the set of constraints are \textit{separable}, namely, the design of the $i$-th unitaries $U_i,\tilde{U_i}$ depends only from the $i$-th input $\ket{\varphi_i}$ and it is completely independent from the other inputs as well as the other unitaries. Indeed, as long as all the separable input qubits are all set to the same state $\ket{\eta}$ (which is reasonable), the condition for deterministically generating GHZ states\footnote{It worthwhile to note that this consideration holds also for Bell and W-like states generation.} reduces to a single constraint regardless of the \textit{dimension} of the state to be generated. Namely, the unitaries acting on the different qubits can be the same. This key feature makes the protocol highly scalable.\\
This \textit{pivotal} separability feature of the necessary and sufficient conditions derived in \eqref{Eq:09} (as well as in \eqref{Eq:07}) allows us to easily address the issue of designing unitaries and input state for generating a GHZ-like state. Indeed, similarly to the bipartite case, by assuming $\ket{\eta \eta \eta}$ as input state, with $\ket{\eta} = \sqrt{\alpha} \ket{0} + \sqrt{1-\alpha} \ket{1}$, as well as $U_i = \sigma_z$ and $\tilde{U_i} = R_y(2\lambda)$ for any $i$, we have that the condition for GHZ-like output state given in \eqref{Eq:07} translates to $\lambda = \frac{\pi}{4}$. \textcolor{black}{This is shown in Figure~\ref{Fig:05} by plotting the GME concurrence (see Appendix.~\ref{appendix.1}), which can characterize the entanglement content of the tri-partite output state $\ket{\psi^{(3)}_{\pm}}$ as a function of the parameters\footnote{\textcolor{black}{The GME concurrence characterizes entanglement in tri-partite states. It is an entanglement monotone, it has a maximum of one in GHZ-like states and a maximum of $0.9$ for W-like states, and it decreases with the decrease of entanglement content in the corresponding state.}}. We can appreciate that the GME concurrence vanishes for $\lambda=0$, and reaches its maximum when $\lambda=\frac{\pi}{4}$. Importantly, we notice that in both ranges of the parameter $\lambda$ given by $\lambda\in]0,\frac{\pi}{4}[$ and $\lambda\in]\frac{\pi}{4},\frac{\pi}{2}[$ we have a non-vanishing probability of generating a GHZ-state with a valuable entanglement content.}

\subsection*{W-like states generation}
\label{Sec:2.4}

\begin{figure*}[t]
\color{black}
    \begin{minipage}[c]{0.38\textwidth}
        \begin{center}
            \includegraphics[width=1\columnwidth]{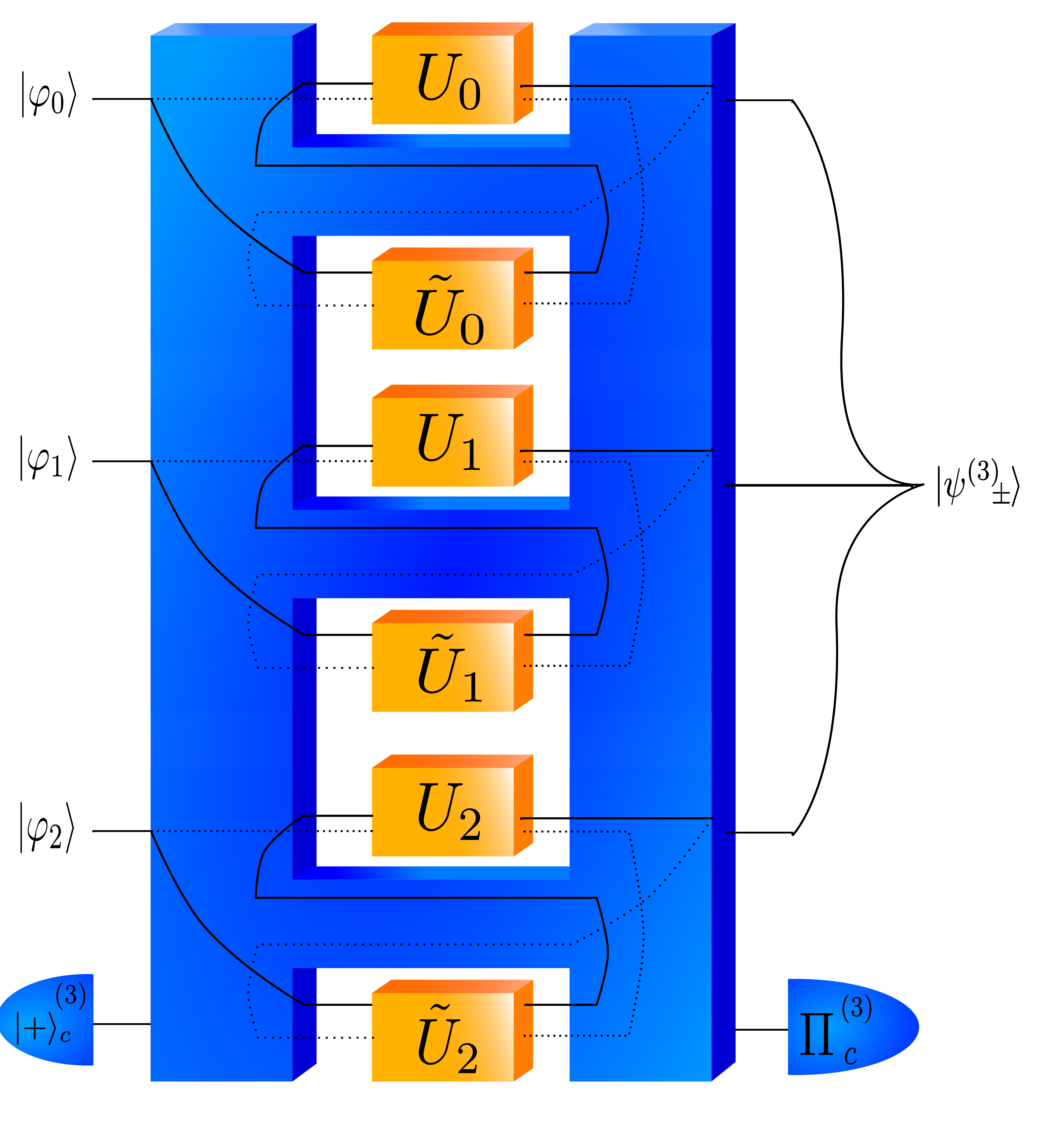}
        \end{center}
        \subcaption{\color{black}Scheme for generating a W-like state through a superposition of the causal order between 1-qubit local unitaries $U_i, \tilde{U_i}$ acting on the $i$-th qubit $\ket{\varphi_i}$.}
        \label{Fig:06-a}
    \end{minipage}
    \hspace{0.03\textwidth}
    \begin{minipage}[c]{0.58\textwidth}
        \begin{center}
            \includegraphics[width=1\columnwidth]{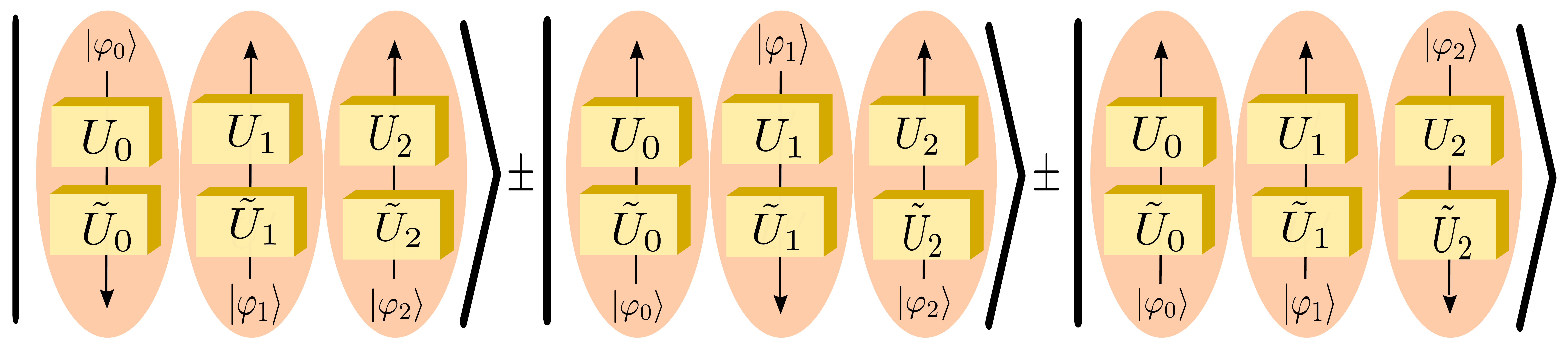}
        \end{center}
        \subcaption{\color{black}Pictorial representation of the output state $\ket{\psi^{(3)}_{\pm \pm}}$, once the control qubit is measured. The output is given by the coherent superposition of three terms, differing for the order between the unitaries acting on each qubit.}  
        \label{Fig:06-b}
    \end{minipage}
    \caption{\color{black}W-like state through a superposition of two alternative causal orders between two 1-qubit local unitaries $U_i$ and $\tilde{U_i}$ operating on the $i$-th qubit of a separable input state $\ket{\varphi_0} \otimes \ket{\varphi_1} \otimes \ket{\varphi_2}$.}
    \label{Fig:06}
    \hrulefill
\end{figure*}

\begin{figure*}
   \begin{minipage}[c]{0.48\textwidth}
        \begin{center}
            \includegraphics[width=1\columnwidth]{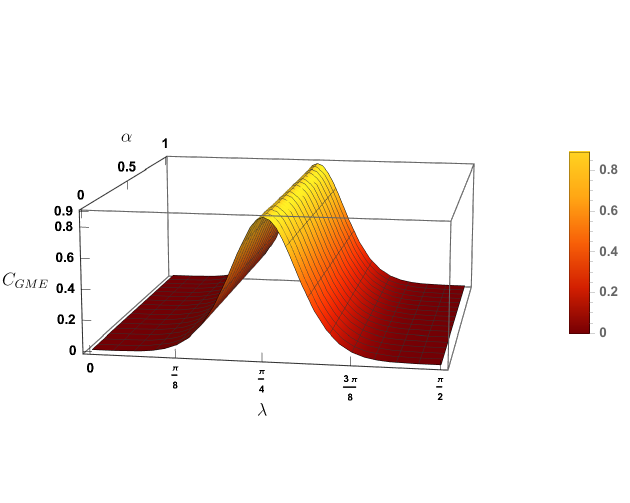}
        \end{center}
        \subcaption{GME concurrence $C_{GME}(\rho_{++})$ of the state $\ket{\psi^{(3)}_{++}}$ given in \eqref{Eq:11}, obtained as output when the control qubit is measured as $\ket{++}$ with $\rho_{++} \eqdef \ket{\psi^{(3)}_{++}}\bra{\psi^{(3)}_{++}}$.}
        \label{Fig:07-a}
    \end{minipage}
    \hspace{0.03\textwidth}
    \begin{minipage}[c]{0.48\textwidth}
        \begin{center}
            \includegraphics[width=1\columnwidth]{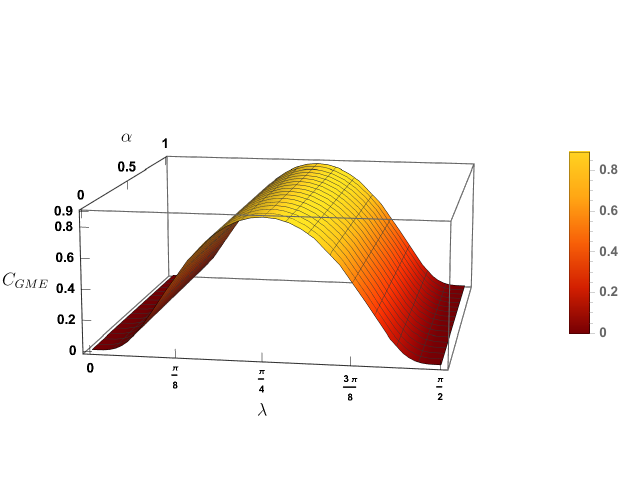}
        \end{center}
        \subcaption{GME concurrence $C_{GME}$ for the states $\ket{\psi^{(3)}_{+-}}$, $\ket{\psi^{(3)}_{--}}$ and $\ket{\psi^{(3)}_{-+}}$ given in \eqref{Eq:11}.}
        \label{Fig:07-b}
    \end{minipage}
    \caption{3D plot for the GME concurrence of the output state $\ket{\psi^{(3)}_{\pm\pm}}$ given in \eqref{Eq:11} as function of: i) the y-rotation parameter $\lambda$ controlling the unitaries $\{ \tilde{U_i} \}$, and ii) the superposition parameter $\alpha$ controlling the input state $\ket{\varphi_0 \varphi_1 \varphi_2}$, assumed real. Unitaries $\{ U_i \}$ set to Pauli-z gate. W-like states exhibit GME concurrence equal to $0.9$.}
    \label{Fig-07}
    \hrulefill
\end{figure*}

To generate W-like states, a coherent control of two alternative evolutions -- as previously done  for both Bell and GHZ-like states -- is not sufficient due to distinguishing peculiarities of W-like states.\\
Basically, the (minimum, for a proper basis choice) number of superpositions grows linearly with the number of parties in the state, in contrast to GHZ states where the (minimum) number of superpositions is two regardless of parties. This is expected, as the W-like states are from a nonequivalent class with respect to GHZ-like states, also by means of stochastic LOCC. For this, to generate W-like states from a coherent superposition of causal orders we need to adjust our coherent control strategy.\\
Let us consider the scheme shown in Figure~\ref{Fig:06-a}, where the $i$-th individual qubit of an initially pure product $3$-partite state $\ket{\varphi_0 \varphi_1 \varphi_2}$ evolves through a superposition of two alternative causal orders between local unitaries $U_i$ and $\tilde{U_i}$ controlled by a degree of freedom $\ket{\varphi_c}$ so that the quantum switch supermap exhibits the unitary Kraus operator $S = \tilde{U_0} U_0 \otimes U_1 \tilde{U_1} \otimes U_2 \tilde{U_2} \otimes \ketbra{0}{0}_c + U_0 \tilde{U_0} \otimes \tilde{U_1} U_1 \otimes U_2 \tilde{U_2} \otimes \ketbra{1}{1}_c + U_0 \tilde{U_0} \otimes U_1 \tilde{U_1} \otimes \tilde{U_2} U_2 \otimes \ketbra{2}{2}_c$. Accordingly, the overall unitary operation on the input state is given by: 
\begin{align}
     S &(\ket{\varphi_0 \varphi_1 \varphi_2}\otimes \ket{\varphi_c})= \nonumber\\
     & \frac{1}{\sqrt{3}}[\left( \tilde{U_0} \otimes U_1 \otimes U_2 \cdot U_0 \otimes \tilde{U_1} \otimes \tilde{U_2} \right)  \ket{\varphi_0 \varphi_1 \varphi_2} \otimes \ket{0}_c +  \nonumber \\
    &  + \left( U_0 \otimes \tilde{U_1} \otimes U_2 \cdot \tilde{U_0} \otimes U_1 \otimes \tilde{U_2} \right) \ket{\varphi_0 \varphi_1 \varphi_1} \otimes \ket{1}_c \nonumber\\
    & + \left( U_0 \otimes U_1 \otimes \tilde{U_2} \cdot \tilde{U_0} \otimes \tilde{U_1} \otimes U_2 \right) \ket{\varphi_1 \varphi_1 \varphi_2} \otimes \ket{2}_c]
	\label{Eq:10}
\end{align}
Upon measuring the control degrees of freedom according to an appropriate measurement setup (see Remark following Theorem~\ref{Theo:03} in Appendix.~\ref{appendix.4}), the following state emerges: 
\begin{align}
    \ket{\psi^{(3)}_{\pm\pm}} =&\frac{1}{\sqrt{L_{\pm\pm}}}
        \Big( \overleftarrow{U_0} \otimes \overrightarrow{U_1} \otimes \overrightarrow{U_2} \nonumber\\
       & \pm \overrightarrow{U_0} \otimes \overleftarrow{U_1} \otimes \overrightarrow{U_2} 
        \pm \overrightarrow{U_0} \otimes \overrightarrow{U_1} \otimes \overleftarrow{U_2} \Big) \ket{\varphi_0 \varphi_1 \varphi_2}
    \label{Eq:11}
\end{align}
with $L_{\pm\pm}$ being the appropriate normalization constant, with $\overleftarrow{\cdot},\overrightarrow{\cdot}$ being the usual shorthand notations for the alternative causal orders among the unitaries, and with $\pm$ being equal to $+$ or $-$ depending on the measurement output of the control qubit. Regardless of the particular expression of $\ket{\psi^{(3)}_{\pm\pm}}$, the state is a superposition of three different input processing, differing for the causal order between the unitaries acting on each qubit. And, regardless of the particular expression, the necessary and sufficient condition for the output in \eqref{Eq:11} being a W-like state is (see Theorem~\ref{Theo:03} in Appendix.~\ref{appendix.4}):
\begin{equation}
	\bra{\varphi_i} \overleftarrow{U_i}^{\dagger} \overrightarrow{U_i} \ket{\varphi_i} = 0 \quad \forall \, i=0,1,2
    \label{Eq:12}
\end{equation}
Furthermore, similarly to the GHZ-like state, there exists a lighter condition for the separability of the output state in \eqref{Eq:11}, given by (see Proposition~\ref{Prop:03} in Appendix.~\ref{appendix.4})
\begin{equation}
\exists i \in \{0,1,2\} : \overleftarrow{U_i} \ket{\varphi}=\overrightarrow{U_i} \ket{\varphi}
\label{Eq:13}
\end{equation}
It is worthwhile to note that the scheme in Figure~\ref{Fig:06} straightforwardly extends to $n$-partite W-like states by simply extending condition \eqref{Eq:12} to any $i=0, \ldots, n-1$, by reasoning as highlighted in the Remark following Theorem.~\ref{Theo:02} in Appendix.~\ref{appendix.3}. Furthermore, the same considerations in terms of unitaries design made for the GHZ-like states continue to hold. This is  confirmed by assuming -- as done for the GHZ-like states -- as input state $\ket{\eta \eta \eta}$, with $\ket{\eta} = \sqrt{\alpha} \ket{0} + \sqrt{1-\alpha} \ket{1}$, as well as $U_i = \sigma_z$ and $\tilde{U_i} = R_y(2\lambda)$ for any $i$. With this setting, the necessary and sufficient condition for the output being a W-like state translates to $\lambda = \frac{\pi}{4}$. This can be clearly seen from the visualization of the GME concurrence of the states in \eqref{Eq:11} given in Fig.~\ref{Fig-07}, where the GME concurrence for all states reaches the maximum at the critical value of $\lambda=\frac{\pi}{4}$, whereas it vanishes for $\lambda=0,\frac{\pi}{2}$ when the output state becomes separable. \textcolor{black}{It is clear from the plots that W-like states generation is different from the previous cases of GHZ-like and Bell states, in the sense that the GME concurrence, namely, the entanglement content, of the output states falls down vary rapidly when moving away from the optimality point $\lambda=\frac{\pi}{4}$.}


\subsection*{Graph states generation}
An important class of resourceful states in many quantum information protocols are graph states. Their generation and distribution in quantum networks would be considered as a fundamental network function. Indeed, graph states will increase the power of communication networks in terms of security and performance. In particular, many communication bottlenecks can be outpassed -- i.e., on demand extraction of multiple EPR pairs -- and measurement-based quantum computing can be achieved.\\
Unlike the cases analyzed in the previous sections, different considerations might be taken into account for the generation of graph states. More into details, depending on different descriptions -- full or partial -- of the considered graph state, one can have different superposition of local causal orders strategies. Indeed, if one has full knowledge of the targeted multipartite entangled state to be generated, a strategy following our previous discussions of GHZ- and W-like states, can be established. Although, this strategy is not optimal in general in terms of the control overhead, whenever the state description is not given in its optimal (canonical) form. Differently, if one is limited to the  knowledge of the graph state in terms of its adjacency matrix, careful tailing of the underlying graph topology is necessary.\\
Hence, the question that we may ask is: can we design an indefinite causal strategy of local unitaries based on the graph topology underlying a targeted graph state? If the adjacency matrix of the targeted graph state is known, we can design a corresponding indefinite causal order of local unitaries acting on single qubit unitaries, which generates a state equivalent to the targeted graph state up to a local unitary. Specifically, let $G=(V,E)$ be the graph corresponding to the targeted graph state $\ket{G}$. The indefinite causal order of local unitaries strategy generating such a state is given by:
\begin{equation}
    S=\displaystyle\otimes_{(i,j) \in E}\Big[V^{(i,j)}\tilde{V}^{(i,j)}\otimes\ketbra{0}{0}_c+\tilde{V}^{(i,j)}V^{(i,j)}\otimes\ketbra{1}{1}_c\Big]
    \label{eq:22}
\end{equation}
with $V^{(i,j)} = U_i \otimes U_j$ and $\tilde{V}^{(i,j)} = \tilde{U}_i \otimes \tilde{U}_j$ denoting local unitary operators acting on qubits $i$ and $j$, and satisfying the constraints given in \eqref{Eq:05}.\\
It is easy to note that the dimension $d_c$ of the control degree of freedom is given by $d_c=2^{|E|}$ with $|E|$ denoting the cardinality of the set of edges $E$. Differently from the previous cases, the single qubit unitaries overhead needed in this process is equal to $4|E|$. Namely, by optimizing the number of edges over a the local Clifford equivalence class of the target state -- i.e, by finding the graph with least number of edges -- we can find an effective sub-optimal control strategy, knowing that the optimal strategy is the one corresponding to the finest Schmidt rank of the state. Indeed, the state generated by the strategy given in \eqref{eq:22} is equivalent to the graph state $\ket{G}$ described by the graph $G=(V,E)$ up to a local unitary, i.e, it belongs to the LU equivalence class of the targeted graph state as detailed in Appendix.~\ref{appendix.5}.\\
Importantly, the previous indefinite causal order strategy in \eqref{eq:22} exploits a composition of many quantum switches in parallel, making it a general operational strategy for all graph states. Nevertheless, one can design non-operational strategies, for specific classes of graph states, which are optimal in terms of the dimension of the control degree of freedom. For instance, one can focus on the class of graph states for which the bipartite maximal entanglement rank $r^*=\max_k r_k$  is equal to the Schmidt rank of their finest cut \cite{hein}. Indeed, this class encompasses bi-colorable cluster states, constituting resourceful states for one-way quantum computing. Namely, if an indefinite causal order strategy of single qubit unitaries $S=\sum_{j=1}^{2^{r^*}}\otimes_{i=1}^{n} U_i^{(j)}\otimes \ketbra{j}{j}_c$ is claimed to be generating a target graph state $\ket{G}=\sum_{j=1}^{2^{r^*}}\otimes_{i=1}^{n} U_i^{(j)}\ket{0}_i$, with entanglement ranks $\{r_k\}_{k=1}^p$ on different bipartitions,  conditions verifying its validity can be designed. These conditions are necessary for the certification of a given $S$ and they are given by:

\begin{strip}
\rule[-1ex]{\columnwidth}{1pt}\rule[-1ex]{1pt}{1.5ex}
\begin{align}
    \label{Eq:17}
    &\bra{\psi}_h|\psi\rangle_{h'}=\Bigg[\sum_{{j'}_{h'}}^{2^{l}}\Big[\otimes_{i\in A}\bra{0}_iU_i^{({j'}_{h'})\dagger}\Big]\Bigg] \Bigg[\sum_{{j}_{h}}^{2^{l}}\Big[\otimes_{i\in A}U_i^{({j}_{h})}\ket{0}_i\Big]\Bigg]=\delta^{hh'}\nonumber\\
   & \bra{\phi}_h|\phi\rangle_{h'}= \Bigg[\sum_{{j'}_{h'}}^{2^{k}}\Big[\otimes_{i\in B}\bra{0}_iU_i^{({j'}_{h'})\dagger}\Big]\Bigg] \Bigg[\sum_{{j}_{h}}^{2^{k}}\Big[\otimes_{i\in B}U_i^{({j}_{h})}\ket{0}_i\Big]\Bigg]=\delta^{hh'}
\end{align}
\hfill\rule[1ex]{1pt}{1.5ex}\rule[2.3ex]{\columnwidth}{1pt}
\end{strip}
where $\ket{G}=\sum_{h=1}^{2^{r'}}\ket{\phi}_h\ket{\psi}_h$ is the schmidt decomposition of the state $\ket{G}$ in the bipartition $\{A,B\}$, and the parameters $r', l, k$ should satisfy $l+k+r'=r^*$ with $r^*$ being the maximum entanglement rank of the state, as is detailed in Appendix.~\ref{appendix.5}.  One can easily verify that the conditions given in \eqref{Eq:09} for $GHZ$ states can be directly derived from the conditions in \eqref{Eq:17}, as $GHZ$ states are a particular instance of graph states. Conversely, the conditions for the generation of $W$ states cannot be derived from the conditions given in \eqref{Eq:17} as $W$ states do not have a graph representation being a class of Dicke states.

\section{Discussion}
\label{Sec:03}
In the previous section, we have shown that the generation of GME states belonging to different non-equivalent classes of states is deterministically achievable through a proper superposition of causal order between local unitaries. We discuss now the possible applications of the proposed scheme under two complementary perspectives, namely, entanglement generation and entanglement distribution, for the Quantum Internet \cite{CacCalTafCatGherBia2020,Pirandola2016UniteTB,Wehner2018QuantumIA}.

\subsection*{Local Entanglement Generation}
\label{Sec:03.1}
We first consider the case where the entanglement generation is \textit{local} rather than \textit{distributed}. Namely, the local unitaries $U_i, \tilde{U}_i$ as well as the controlling degree of freedom $\ket{\varphi_c}$ are all located within the same quantum node, which locally implements the proper supermap \textcolor{black}{-- such as \eqref{Eq:06} illustrated in  Figure~\ref{Fig:01-a} with many-partite local single qubit unitaries. or \eqref{Eq:10} illustrated in Figure~\ref{Fig:06-a} --} for generating GME states.\\
In this scenario, the entangled states are thus deterministically generated in a (some) network node -- acting as entanglement generator -- and they are subsequently distributed within the network through proper quantum communication links. \\
Hence, the network node acting as entanglement generator implements a coherent control strategy on the local unitaries processing some initial product state -- with both the unitaries and the input state considered as free resources -- for generating the GME states. Clearly, as discussed within the paper, the coherent control strategy depends on the desired GME output state. Although it is still technologically unclear whether a node can dynamically change the coherent control strategy for generating entangled states belonging to different non-equivalent classes, the proposed scheme for deterministic entanglement generation is achievable in near-term quantum networks, as the coherent control of orders of operations is affordable by current technology level and it has been successfully implemented for single-qubit channels \cite{Procopio_2015,Rubino_2017,8,white,WeiJian}.\\
Regardless the control strategy being dynamic or fixed a-priori, three are the crucial properties of the proposed scheme for deterministic GME state generation. i) First, the individual input qubits don't interact each others or with the control in any way. Indeed, they only traverse their respective local unitaries $U_, \tilde{U_i}$ in a coherent superposition of two alternative causal orders. ii) Second, the (sufficient and necessary) condition for deterministically generating GHZ- and W-like states consists of separable constraints. Namely, the design of the $i$-th unitaries is completely independent from the other unitaries as well as any input qubit different from $\ket{\varphi_i}$. iii) Third, as long as all the separable input qubits are all set to the same state $\ket{\eta}$ (which is reasonable), the condition for deterministically generating GHZ- and W-like states reduces to a single constraint regardless of the \textit{dimension} of the state to be generated. Namely, the unitaries acting on the different qubits can be the same. These crucial features make the protocol highly scalable.
\textcolor{black}{These features make the proposed framework an ideal candidate for the design of multi-qubit gates on photonic platforms. Interestingly, this design would be achieved through single-qubit-only gates. In contrast, building multi-qubit photonic gates is usually a hard task to achieve as photons barely interact with each other. This stands as a major obstacle for the design of all-photonic quantum computing.}
    
\subsection*{Distributed Entanglement Generation}

\begin{figure*}[t]
    \begin{minipage}[c]{0.48\textwidth}
        \begin{center}
            \includegraphics[width=1\columnwidth]{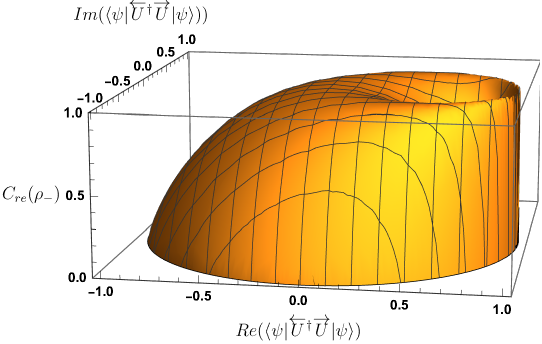}
        \end{center}
        \subcaption{Scheme for distributing multipartite entanglement states in a quantum network. The network is composed by nine clients, organized in three sets\textcolor{black}{,} each served by an \textit{intermediate entangler} whose control qubit is controlled by a \textit{entanglement coordinator}.}
         \label{Fig:09-b}
    \end{minipage}
    \hspace{0.03\textwidth}
    \begin{minipage}[c]{0.48\textwidth}
        \begin{center}
            \includegraphics[width=1\textwidth]{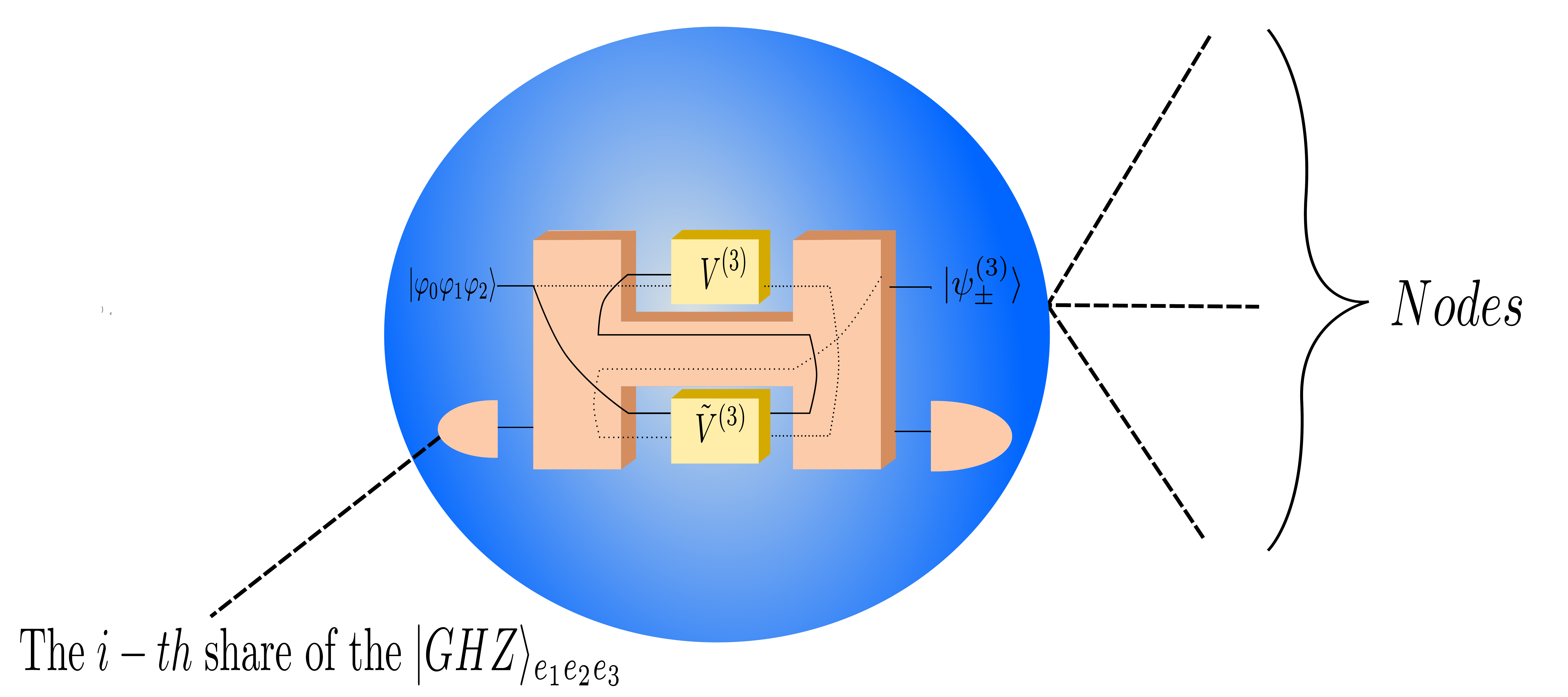}
        \end{center}
        \subcaption{\color{black}A magnification of the intermediate entangler $e_i$, which exploits a superposition of causal orders to distribute multi-partite entanglement to remote nodes.}
        \label{Fig:09-c}
    \end{minipage}
    \caption{Distributed multi-partite generation.}
    \label{Fig:09}
    \hrulefill
\end{figure*}

When it comes to multipartite entanglement distribution in a quantum network, several issues arises. First, as the number of parties to be entangled increases, the number of required multi-qubit gates increases as well. This not only implies severe error propagation effects, but it also hardly scales -- as instance, in W-like states -- with the number of parties. Furthermore, regardless of the number of parties, whenever the size of the quantum network grows to moderate- or large-scale \cite{PirRicOtt_17}, direct entanglement distribution is not feasible anymore due to photon noise and losses. In this context, \textit{quantum repeaters}\cite{BriegelDur,Wehner2018QuantumIA,Rodney} are commonly accepted as the strategy for increasing the entanglement distribution range. Unfortunately, regardless of the repeater particulars that roughly depend on the repeater \textit{generation}\cite{Muralidharan2016OptimalAF}, quantum repeaters require some sort of Bell state measurements for Bell pairs distribution or other projective measurements for  multipartite entangled states, which are usually hard to implement and very noisy in practice.\\
Interestingly, our scheme for entanglement generation could provide an alternative strategy for overcoming such issues in distributing multi-partite entanglement, without the need of multi-qubits gates or any other interaction among the qubits. Let us better clarify this with an example. Specifically, as shown in Figure~\ref{Fig:09-b}, multiple quantum switches -- referred to as \textit{edge entanglers} -- are geographically distributed through the network, so that each switch is closely located (from the entanglement distribution perspective) to a certain group of nodes. Each edge entangler implements the proper supermap -states\textcolor{black}{ - such as \eqref{Eq:06} illustrated in Figure~\ref{Fig:01-a} for GHZ-like when many-parties local qubit unitaries are considered-} for generating GME states. But each edge entangler uses, as control degree of freedom, the output of another quantum switch -- which acts as \textit{entangler coordinator} -- in order to collectively generate the desired multi-partite entangled state. Clearly, each edge entangler generates the required $k$-partite GHZ-like state according to the number $k$ of nodes that are physically linked to it. As instance, in Figure~\ref{Fig:09-b}, $e_1$, $e_2$ and $e_3$ generate a tri-partite entangled state by relying on the coherent control of the causal order between unitaries $V^{3}$ and $\tilde{V}^{3}$ as illustrated in Figure~\ref{Fig:09-c}, where the coherent control is provided as a $3$-GHZ state generated by the entanglement coordinator $e_0$. The overall state distributed through the network is a valid $\ket{GHZ}_9$ state. It is worthwhile to note that the proposed scheme can scale to large networks through a proper hierarchical multi-tier architecture, where additional \textit{intermediate entanglers} are deployed between the coordinator and the edge entanglers.\\
It is important to note that, in the above example, we have only discussed the distributed multipartite entanglement generation when the coherent control of remotely located edge entanglers is obtained through a proper multi-partite entangled state, shared between the edge entanglers. However, the scheme in Figure~\ref{Fig:09-b} requires only the availability of a coherent control of the unitaries among remotely located edge entanglers, regardless of the specific implementation of such a control. And such a coherent control is considered the \textit{genuine quantum feature} of a quantum network\cite{duer-2019}, where a genuine quantum coherence is an intrinsic property of the communication network.

It is worth noting that by adopting an indefinite causal order generating $GHZ$-like states in each entangler would only establish a $GHZ$-like state between the client nodes. Conversely, an interesting feature of our scheme is the ability to distributing graph states in the network. In fact, these states constitutes the fundamental resource for measurement-based quantum computing, and their distribution in future quantum networks plays an essential role.\\
In the following, let us provide an example of graph state distribution, by following a scheme similar to Figure~\ref{Fig:09-b}.
Specifically, the entangler coordinator $e_0$ generates a Bell state, which is distributed to the two intermediate entanglers $e_1$ and $e_2$ as control degree of freedom. We refer to these controls as $\varphi_c^1$ and $\varphi_c^2$, respectively. Indeed, an additional control degree of freedom is required at one of the intermediate entangles -- says $e_2$-- and it is referred to as $\varphi_{\tilde{c}}^2$ and initialized in $\ket{+}$. By denoting with $\varphi^i_0$ and $\varphi^i_1$ the two inputs to the $i$-th intermediate entangler, the overall global state is given by:
\begin{align}
    &\frac{1}{\sqrt{2}}\ket{0}_{\varphi_c^1}\ket{0}_{\varphi_c^2}\ket{00}_{\varphi^1_0,\varphi^1_1}\ket{+}_{\varphi_{\tilde{c}}^2}\ket{00}_{\varphi^2_0,\varphi^2_1}\nonumber\\
    &+ \frac{1}{\sqrt{2}}\ket{1}_{\varphi_c^1}\ket{1}_{\varphi_c^2}\ket{00}_{\varphi^1_0,\varphi^1_1}\ket{+}_{\varphi_{\tilde{c}}^2}\ket{00}_{\varphi^2_0,\varphi^2_1}
\end{align}
The intermediate entangler $e_1$ uses its share $\varphi_c^1$ as a control of indefinite causal order process to establish entanglement between the inputs ${\varphi^1_0}$ and ${\varphi^1_1}$. Similarly, the intermediate entangler $e_2$ uses its share $\varphi_c^2$ along with $\varphi_{\tilde{c}}^2$ as a two-qubit control degree of freedom of the appropriate indefinite causal order process, to create entanglement between the inputs ${\varphi^2_0}$ and ${\varphi^2_1}$. After separable  measurement on the controls in the appropriate maximally coherent bases, the following graph state is obtained:
\begin{align}
    &\frac{1}{2}\ket{00}_{\varphi^1_0,\varphi^1_1}(\ket{00}\pm\ket{11})_{\varphi^2_0,\varphi^2_1}+\frac{1}{2}\ket{11}_{\varphi^1_0,\varphi^1_1}(\ket{10}\pm\ket{01})_{\varphi^2_0,\varphi^2_1}
\end{align}
Upon the sequential distribution of $\varphi^1_0,\varphi^1_1$ and $\varphi^2_0,\varphi^2_1$ to the corresponding clients $n_1$,$n_2$,$n_3$ and $n_4$ respectively, the above emerging graph state is deterministically distributed. We should note that the same strategy establishes linear and ring four-partite cluster states.\\
The proposed protocol for distributed entanglement generation opens a new research direction based on the advantage of the indefinite causal order framework for future communication networks. Besides, the applicability of such a protocol relies on the ability of distributing pure entangled states which might be achieved by different multipartite entanglement distillation protocols \cite{Dur-2019}. Nevertheless, careful investigation of the robustness and resilience of the proposed scheme to noise would have significant importance for the building of fault tolerant quantum communication networks. These considerations might be treated in different layers of a quantum protocol stack \cite{IllCalMan-22}, aiming at harnessing the full quantum potential of future quantum networks in an efficient way.

\color{black}
\subsection*{Entanglement mapping}

\begin{figure}[t]
\color{black}
    \begin{center}
        \includegraphics[width=0.7\columnwidth]{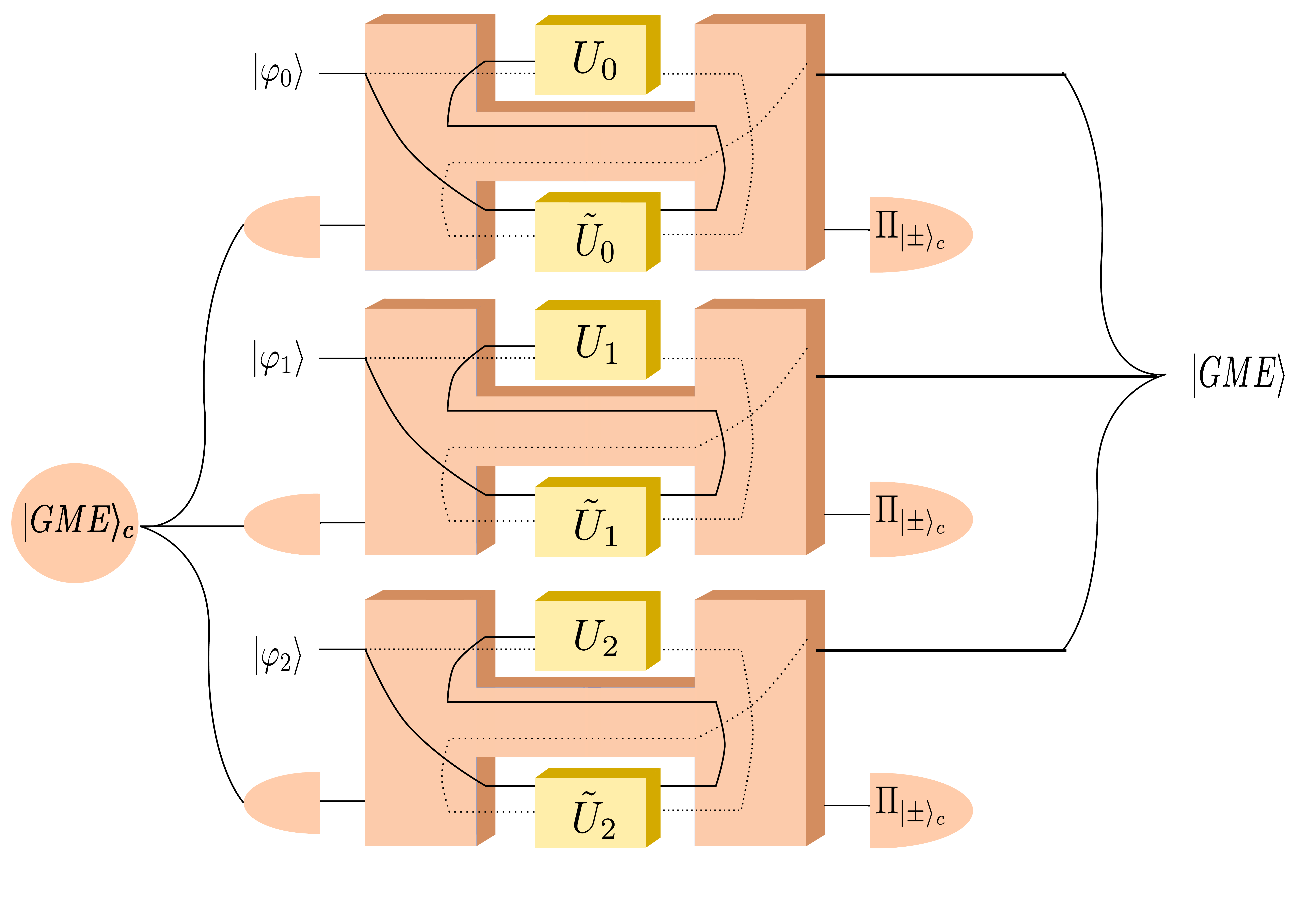}
    \end{center}
    \caption{\color{black}Scheme for entanglement mapping between different quantum degrees of freedom. By implementing a coherent control of three different quantum switches through a proper 3-partite GME state and by measuring each control qubit in the coherent basis, an initially pure product tripartite state $\ket{\varphi_0 \varphi_1 \varphi_2}$ is deterministically transformed into a 3-partite GME state.} 
    \label{Fig:08}
     \hrulefill
\end{figure}

Here we consider a scenario where entanglement -- rather then generated -- must be mapped between different quantum degrees of freedom.
As an example, let us consider deterministic generation of photonic GME states, which may benefits from a matter-photonic interface with a quantum degree of freedom -- such as superconducting-circuit based qubits \cite{Besse_2020} -- where entanglement can be generated easier than in photonic-circuits. Another example is represented by matter-flying interfaces \textit{per-se}, which represent a critical component for quantum networks \cite{CacCalTafCatGherBia2020, interface,interface2,Interface3}, where matter qubits for information processing/storing -- based on heterogeneous technologies ranging from transmons through  quantum dots to ion traps -- must be interfaced to flying qubits -- generally implemented with photons -- acting as information carriers.\\
Regardless of the specific applications for an entanglement mapper, the proposed scheme based on superposition of causal orders provides an interesting approach toward deterministic entanglement mapping, worthwhile of further investigation. As an example, let us consider the scheme shown in Figure~\ref{Fig:08}. The initially-entangled quantum degree of freedom, say\footnote{Clearly, the proposed scheme applies to the deterministic mapping of $W$-like states as well.} in a GHZ state, is used to implement a coherent control among different quantum switches. Each quantum switch implements the superposition of causal orders between two unitaries given in \eqref{Eq:03} and shown in Figure~\ref{Fig:01-a}, by acting on the individual qubit $\ket{\varphi_i}$ of a second quantum degree of freedom, initially in a separable state. As long as the condition for deterministic generation of GHZ-like state in \eqref{Eq:07} is satisfied, the GHZ state is deterministically mapped from the control degree of freedom to the initially separable second degree of freedom, which becomes maximally entangled. In a nutshell, the scheme harnesses the quantum correlation embedded within the control degree of freedom to generate -- through multiple switch instances -- a coherent evolution of the input degree of freedom, which eventually exhibits a correlation of the same nature (i.e., class) of the original entanglement.\\
It is worthwhile to note that the proposed mapping scheme does not require any interaction between the input qubits or between the input qubit and the control qubit, which represents a key feature whenever the input qubits weakly interact each others or with the environment, as in the mentioned case of photonic qubits. \textcolor{black}{We should point that the entanglement mapping from the control degrees of freedom of quantum switches placed in parallel to the corresponding target states has been independently studied in a different framework \cite{Rubino-2022}.}

\color{black}

\color{black}


\color{black}

\bibliographystyle{IEEEtran}
\bibliography{ref}

\begin{thebibliography}{10}
\providecommand{\url}[1]{#1}
\csname url@samestyle\endcsname
\providecommand{\newblock}{\relax}
\providecommand{\bibinfo}[2]{#2}
\providecommand{\BIBentrySTDinterwordspacing}{\spaceskip=0pt\relax}
\providecommand{\BIBentryALTinterwordstretchfactor}{4}
\providecommand{\BIBentryALTinterwordspacing}{\spaceskip=\fontdimen2\font plus
\BIBentryALTinterwordstretchfactor\fontdimen3\font minus
  \fontdimen4\font\relax}
\providecommand{\BIBforeignlanguage}[2]{{%
\expandafter\ifx\csname l@#1\endcsname\relax
\typeout{** WARNING: IEEEtran.bst: No hyphenation pattern has been}%
\typeout{** loaded for the language `#1'. Using the pattern for}%
\typeout{** the default language instead.}%
\else
\language=\csname l@#1\endcsname
\fi
#2}}
\providecommand{\BIBdecl}{\relax}
\BIBdecl

\bibitem{CacCalTafCatGherBia2020}
A.~S. Cacciapuoti, M.~Caleffi, F.~Tafuri, F.~S. Cataliotti, S.~Gherardini, and
  G.~Bianchi, ``Quantum internet: Networking challenges in distributed quantum
  computing,'' \emph{IEEE Network}, vol.~34, no.~1, pp. 137--143, 2020.

\bibitem{IllCalMan-22}
J.~Illiano, M.~Caleffi, A.~Manzalini, and A.~S. Cacciapuoti,
  ``{\textcolor{black} Quantum Internet protocol stack: A comprehensive
  survey},'' \emph{Computer Networks}, p. 109092, 2022.

\bibitem{Pirandola2016UniteTB}
S.~Pirandola and S.~Braunstein, ``Physics: Unite to build a quantum internet,''
  \emph{Nature}, vol. 532, pp. 169--171, 2016.

\bibitem{Wehner2018QuantumIA}
S.~Wehner, D.~Elkouss, and R.~Hanson, ``Quantum internet: A vision for the road
  ahead,'' \emph{Science}, vol. 362, 2018.

\bibitem{Wilde}
M.~M. Wilde, \emph{Quantum Information Theory}, 2nd~ed.\hskip 1em plus 0.5em
  minus 0.4em\relax Cambridge University Press, 2017.

\bibitem{nielsen_chuang_2010}
M.~A. Nielsen and I.~L. Chuang, \emph{Quantum Computation and Quantum
  Information: 10th Anniversary Edition}.\hskip 1em plus 0.5em minus
  0.4em\relax Cambridge University Press, 2010.

\bibitem{Flhmann2019EncodingAQ}
C.~Fl{\"u}hmann, T.~L. Nguyen, M.~Marinelli, V.~Negnevitsky, K.~K. Mehta, and
  J.~P. Home, ``Encoding a qubit in a trapped-ion mechanical oscillator,''
  \emph{Nature}, vol. 566, pp. 513--517, 2019.

\bibitem{arute2019quantum}
F.~Arute, K.~Arya, R.~Babbush, D.~Bacon, J.~C. Bardin, R.~Barends, R.~Biswas,
  S.~Boixo, F.~G. Brandao, D.~A. Buell \emph{et~al.}, ``Quantum supremacy using
  a programmable superconducting processor,'' \emph{Nature}, vol. 574, no.
  7779, pp. 505--510, 2019.

\bibitem{boixo2018characterizing}
S.~Boixo, S.~V. Isakov, V.~N. Smelyanskiy, R.~Babbush, N.~Ding, Z.~Jiang, M.~J.
  Bremner, J.~M. Martinis, and H.~Neven, ``Characterizing quantum supremacy in
  near-term devices,'' \emph{Nature Physics}, vol.~14, no.~6, pp. 595--600,
  2018.

\bibitem{lund2017quantum}
A.~P. Lund, M.~J. Bremner, and T.~C. Ralph, ``Quantum sampling problems,
  bosonsampling and quantum supremacy,'' \emph{npj Quantum Information},
  vol.~3, no.~1, pp. 1--8, 2017.

\bibitem{WanCheLuo_2016}
\BIBentryALTinterwordspacing
X.-L. Wang, L.-K. Chen, W.~Li, H.-L. Huang, C.~Liu, C.~Chen, Y.-H. Luo, Z.-E.
  Su, D.~Wu, Z.-D. Li, and et~al., ``Experimental ten-photon entanglement,''
  \emph{Physical Review Letters}, vol. 117, no.~21, Nov 2016. [Online].
  Available: \url{http://dx.doi.org/10.1103/PhysRevLett.117.210502}
\BIBentrySTDinterwordspacing

\bibitem{RosAraFabTre_2013}
\BIBentryALTinterwordspacing
J.~Roslund, R.~M. de~Araújo, S.~Jiang, C.~Fabre, and N.~Treps,
  ``Wavelength-multiplexed quantum networks with ultrafast frequency combs,''
  \emph{Nature Photonics}, vol.~8, no.~2, p. 109–112, Dec 2013. [Online].
  Available: \url{http://dx.doi.org/10.1038/nphoton.2013.340}
\BIBentrySTDinterwordspacing

\bibitem{CaiRosFerArz_2017}
Y.~Cai, J.~Roslund, G.~Ferrini, F.~Arzani, X.~Xu, C.~Fabre, and N.~Treps,
  ``Multimode entanglement in reconfigurable graph states using optical
  frequency combs,'' \emph{Nature Communications}, vol.~8, 2017.

\bibitem{Besse_2020}
\BIBentryALTinterwordspacing
J.-C. Besse, K.~Reuer, M.~C. Collodo, A.~Wulff, L.~Wernli, A.~Copetudo,
  D.~Malz, P.~Magnard, A.~Akin, M.~Gabureac, and et~al., ``Realizing a
  deterministic source of multipartite-entangled photonic qubits,''
  \emph{Nature Communications}, vol.~11, no.~1, Sep 2020. [Online]. Available:
  \url{http://dx.doi.org/10.1038/s41467-020-18635-x}
\BIBentrySTDinterwordspacing

\bibitem{Schwartz_2016}
\BIBentryALTinterwordspacing
I.~Schwartz, D.~Cogan, E.~R. Schmidgall, Y.~Don, L.~Gantz, O.~Kenneth, N.~H.
  Lindner, and D.~Gershoni, ``Deterministic generation of a cluster state of
  entangled photons,'' \emph{Science}, vol. 354, no. 6311, p. 434–437, Sep
  2016. [Online]. Available: \url{http://dx.doi.org/10.1126/science.aah4758}
\BIBentrySTDinterwordspacing

\bibitem{Istrati_2020}
\BIBentryALTinterwordspacing
D.~Istrati, Y.~Pilnyak, J.~C. Loredo, C.~Antón, N.~Somaschi, P.~Hilaire,
  H.~Ollivier, M.~Esmann, L.~Cohen, L.~Vidro, and et~al., ``Sequential
  generation of linear cluster states from a single photon emitter,''
  \emph{Nature Communications}, vol.~11, no.~1, Oct 2020. [Online]. Available:
  \url{http://dx.doi.org/10.1038/s41467-020-19341-4}
\BIBentrySTDinterwordspacing

\bibitem{Takeda_2019}
\BIBentryALTinterwordspacing
S.~Takeda, K.~Takase, and A.~Furusawa, ``On-demand photonic entanglement
  synthesizer,'' \emph{Science Advances}, vol.~5, no.~5, p. eaaw4530, May 2019.
  [Online]. Available: \url{http://dx.doi.org/10.1126/sciadv.aaw4530}
\BIBentrySTDinterwordspacing

\bibitem{3}
G.~Chiribella and H.~Kristjánsson, ``Quantum shannon theory with
  superpositions of trajectories,'' \emph{Proceedings of the Royal Society A:
  Mathematical, Physical and Engineering Sciences}, vol. 475, no. 2225, p.
  20180903, May 2019.

\bibitem{8}
G.~Rubino, L.~A. Rozema, D.~Ebler, H.~Kristjánsson, S.~Salek,
  P.~Allard~Guérin, A.~A. Abbott, C.~Branciard, C.~Brukner, G.~Chiribella, and
  et~al., ``Experimental quantum communication enhancement by superposing
  trajectories,'' \emph{Physical Review Research}, vol.~3, no.~1, Jan 2021.

\bibitem{10}
G.~{Chiribella}, M.~{Wilson}, and H.~F. {Chau}, ``{Quantum and Classical Data
  Transmission Through Completely Depolarising Channels in a Superposition of
  Cyclic Orders},'' \emph{arXiv e-prints}, p. arXiv:2005.00618, May 2020.

\bibitem{13}
H.~Kristjánsson, G.~Chiribella, S.~Salek, D.~Ebler, and M.~Wilson, ``Resource
  theories of communication,'' \emph{New Journal of Physics}, vol.~22, no.~7,
  p. 073014, Jul 2020.

\bibitem{18}
D.~Ebler, S.~Salek, and G.~Chiribella, ``Enhanced communication with the
  assistance of indefinite causal order,'' \emph{Physical Review Letters}, vol.
  120, no.~12, Mar 2018.

\bibitem{19}
G.~Chiribella, M.~Banik, S.~S. Bhattacharya, T.~Guha, M.~Alimuddin, A.~Roy,
  S.~Saha, S.~Agrawal, and G.~Kar, ``Indefinite causal order enables perfect
  quantum communication with zero capacity channels,'' \emph{New Journal of
  Physics}, Feb 2021.

\bibitem{20}
S.~{Salek}, D.~{Ebler}, and G.~{Chiribella}, ``{Quantum communication in a
  superposition of causal orders},'' \emph{arXiv e-prints}, p.
  arXiv:1809.06655, Sep. 2018.

\bibitem{37}
G.~Chiribella, G.~M. D’Ariano, and P.~Perinotti, ``Transforming quantum
  operations: Quantum supermaps,'' \emph{EPL (Europhysics Letters)}, vol.~83,
  no.~3, p. 30004, Jul 2008.

\bibitem{38}
------, ``Theoretical framework for quantum networks,'' \emph{Physical Review
  A}, vol.~80, no.~2, Aug 2009.

\bibitem{83}
G.~Chiribella, G.~M. D'Ariano, P.~Perinotti, and B.~Valiron, ``Quantum
  computations without definite causal structure,'' \emph{Phys. Rev. A},
  vol.~88, p. 022318, Aug 2013.

\bibitem{Rubino_2017}
G.~Rubino, L.~A. Rozema, A.~Feix, M.~Araújo, J.~M. Zeuner, L.~M. Procopio,
  C.~Brukner, and P.~Walther, ``Experimental verification of an indefinite
  causal order,'' \emph{Science Advances}, vol.~3, no.~3, p. e1602589, Mar
  2017.

\bibitem{Procopio_2015}
L.~M. Procopio, A.~Moqanaki, M.~Araújo, F.~Costa, I.~Alonso~Calafell, E.~G.
  Dowd, D.~R. Hamel, L.~A. Rozema, C.~Brukner, and P.~Walther, ``Experimental
  superposition of orders of quantum gates,'' \emph{Nature Communications},
  vol.~6, no.~1, Aug 2015.

\bibitem{procopio-2020}
L.~M. Procopio, F.~Delgado, M.~Enr\'{\i}quez, N.~Belabas, and J.~A. Levenson,
  ``Sending classical information via three noisy channels in superposition of
  causal orders,'' \emph{Phys. Rev. A}, vol. 101, p. 012346, Jan 2020.

\bibitem{40}
E.~Castro-Ruiz, F.~Giacomini, and C.~Brukner, ``Dynamics of quantum causal
  structures,'' \emph{Physical Review X}, vol.~8, no.~1, Mar 2018.

\bibitem{koudia-2021}
S.~Koudia, A.~S. Cacciapuoti, and M.~Caleffi, ``How deep the theory of quantum
  communications goes: Superadditivity, superactivation and causal
  activation,'' 2021.

\bibitem{koudia2021quantum}
S.~Koudia and A.~Gharbi, ``Quantum non-gaussianity from an indefnite causal
  order of gaussian operations,'' \emph{International Journal of Quantum
  Information}, p. 2150026, 2021.

\bibitem{koudia2019superposition}
------, ``Superposition of causal orders for quantum discrimination of quantum
  processes,'' \emph{International Journal of Quantum Information}, vol.~17,
  no.~07, p. 1950055, 2019.

\bibitem{Miao}
W.~Yokojima, M.~T. Quintino, A.~Soeda, and M.~Murao, ``Consequences of
  preserving reversibility in quantum superchannels,'' \emph{Quantum}, vol.~5,
  p. 441, 2021.

\bibitem{white}
\BIBentryALTinterwordspacing
K.~Goswami, C.~Giarmatzi, M.~Kewming, F.~Costa, C.~Branciard, J.~Romero, and
  A.~G. White, ``Indefinite causal order in a quantum switch,'' \emph{Phys.
  Rev. Lett.}, vol. 121, p. 090503, Aug 2018. [Online]. Available:
  \url{https://link.aps.org/doi/10.1103/PhysRevLett.121.090503}
\BIBentrySTDinterwordspacing

\bibitem{WeiJian}
\BIBentryALTinterwordspacing
K.~Wei, N.~Tischler, S.-R. Zhao, Y.-H. Li, J.~M. Arrazola, Y.~Liu, W.~Zhang,
  H.~Li, L.~You, Z.~Wang, Y.-A. Chen, B.~C. Sanders, Q.~Zhang, G.~J. Pryde,
  F.~Xu, and J.-W. Pan, ``Experimental quantum switching for exponentially
  superior quantum communication complexity,'' \emph{Phys. Rev. Lett.}, vol.
  122, p. 120504, Mar 2019. [Online]. Available:
  \url{https://link.aps.org/doi/10.1103/PhysRevLett.122.120504}
\BIBentrySTDinterwordspacing

\bibitem{guha2022quantum}
{\textcolor{black}Guha, Tamal and Roy, Saptarshi and Chiribella, Giulio},
  ``{\textcolor{black}Quantum networks boosted by entanglement with a control
  system},'' \emph{arXiv preprint arXiv:2206.05247}, 2022.

\bibitem{chiribella_2020}
{\textcolor{black} Hlèr Kristjànsson and Giulio Chiribella and Sina Salek and
  Daniel Ebler and Matthew Wilson}, ``{\textcolor{black}Resource theories of
  communication},'' \emph{New Journal of Physics}, vol.~22, no.~7, p. 073014,
  jul 2020.

\bibitem{Milz-2021}
{\textcolor{black} Milz, Simon and Bavaresco, Jessica and Chiribella, Giulio},
  ``{\textcolor{black} Resource theory of causal connection},'' 2021.

\bibitem{Hein-2006}
{\textcolor{black}Hein, Marc and D{\"u}r, Wolfgang and Eisert, Jens and
  Raussendorf, Robert and Nest, M and Briegel, H-J},
  ``{\textcolor{black}Entanglement in graph states and its applications},''
  \emph{arXiv preprint quant-ph/0602096}, 2006.

\bibitem{ColDarFac-12}
T.~Colnaghi, G.~M. D'Ariano, S.~Facchini, and P.~Perinotti, ``Quantum
  computation with programmable connections between gates,'' \emph{Physics
  Letters A}, vol. 376, no.~45, pp. 2940 -- 2943, 2012.

\bibitem{AraCosBru-14}
M.~Ara\'ujo, F.~Costa, and {\v C}.~Brukner, ``{Computational Advantage from
  Quantum-Controlled Ordering of Gates},'' \emph{Phys. Rev. Lett.}, vol. 113,
  p. 250402, Dec. 2014.

\bibitem{Chi-12}
G.~Chiribella, ``Perfect discrimination of no-signalling channels via quantum
  superposition of causal structures,'' \emph{Phys. Rev. A}, vol.~86, p.
  040301, Oct. 2012.

\bibitem{WakSoeMur-19}
E.~Wakakuwa, A.~Soeda, and M.~Murao, ``{Complexity of Causal Order Structure in
  Distributed Quantum Information Processing: More Rounds of Classical
  Communication Reduce Entanglement Cost},'' \emph{Phys. Rev. Lett.}, vol. 122,
  p. 190502, May 2019.

\bibitem{OreCosBru-12}
O.~Oreshkov, F.~Costa, and {\v C}.~Brukner, ``Quantum correlations with no
  causal order,'' \emph{Nature Communications}, vol.~3, pp. 1092 EP --, Oct.
  2012.

\bibitem{FeiAraBru-15}
A.~Feix, M.~Ara\'ujo, and {\v C}.~Brukner, ``Quantum superposition of the order
  of parties as a communication resource,'' \emph{Phys. Rev. A}, vol.~92, p.
  052326, Nov. 2015.

\bibitem{GueFeiAra-16}
P.~A. Gu\'erin, A.~Feix, M.~Ara\'ujo, and {\v C}.~Brukner, ``{Exponential
  Communication Complexity Advantage from Quantum Superposition of the
  Direction of Communication},'' \emph{Phys. Rev. Lett.}, vol. 117, p. 100502,
  Sep 2016.

\bibitem{cacciapuoti2019capacity}
A.~S. {Cacciapuoti} and M.~{Caleffi}, ``{Capacity Bounds for Quantum
  Communications through Quantum Trajectories},'' \emph{arXiv e-prints}, p.
  arXiv:1912.08575, Dec. 2019.

\bibitem{marcello}
M.~Caleffi and A.~S. Cacciapuoti, ``Quantum switch for the quantum internet:
  Noiseless communications through noisy channels,'' \emph{IEEE Journal on
  Selected Areas in Communications}, vol.~38, no.~3, pp. 575--588, 2020.

\bibitem{CalCacBia2018}
M.~Caleffi, A.~S. Cacciapuoti, and G.~Bianchi, ``Quantum internet: From
  communication to distributed computing!'' in \emph{Proceedings of the 5th ACM
  International Conference on Nanoscale Computing and Communication}, ser.
  NANOCOM '18.\hskip 1em plus 0.5em minus 0.4em\relax New York, NY, USA:
  Association for Computing Machinery, 2018.

\bibitem{daryus}
D.~Chandra, M.~Caleffi, and A.~S. Cacciapuoti, ``The entanglement-assisted
  communication capacity over quantum trajectories,'' \emph{IEEE Transactions
  on Wireless Communications}, pp. 1--1, 2021.

\bibitem{daryus0}
D.~Chandra, A.~S. Cacciapuoti, M.~Caleffi, and L.~Hanzo, ``Direct quantum
  communications in the presence of realistic noisy entanglement,'' \emph{IEEE
  Transactions on Communications}, pp. 1--1, 2021.

\bibitem{39}
O.~Oreshkov, F.~Costa, and C.~Brukner, ``Quantum correlations with no causal
  order,'' \emph{Nature Communications}, vol.~3, no.~1, Jan 2012.

\bibitem{Daniele-2021}
D.~Cuomo, M.~Caleffi, and A.~S. Cacciapuoti, ``Experiencing the communication
  advantage of the superposition of causal orders,'' in \emph{2021 IEEE 22nd
  International Workshop on Signal Processing Advances in Wireless
  Communications (SPAWC)}, 2021, pp. 181--185.

\bibitem{Horodecki_2009}
\BIBentryALTinterwordspacing
R.~Horodecki, P.~Horodecki, M.~Horodecki, and K.~Horodecki, ``Quantum
  entanglement,'' \emph{Reviews of Modern Physics}, vol.~81, no.~2, p.
  865–942, Jun 2009. [Online]. Available:
  \url{http://dx.doi.org/10.1103/RevModPhys.81.865}
\BIBentrySTDinterwordspacing

\bibitem{zych-2019}
\BIBentryALTinterwordspacing
{\textcolor{black}Zych, Magdalena and Costa, Fabio and Pikovski, Igor and
  Brukner, Caslav}, ``{\textcolor{black}Bell’s theorem for temporal order},''
  \emph{Nature Communications}, vol.~10, no.~1, Aug 2019. [Online]. Available:
  \url{http://dx.doi.org/10.1038/s41467-019-11579-x}
\BIBentrySTDinterwordspacing

\bibitem{Rubino-2022}
\BIBentryALTinterwordspacing
{\textcolor{black}Rubino, Giulia and Rozema, Lee A. and Massa, Francesco and
  Araújo, Mateus and Zych, Magdalena and Brukner, Časlav and Walther,
  Philip}, ``{\textcolor{black}Experimental entanglement of temporal order},''
  \emph{Quantum}, vol.~6, p. 621, Jan 2022. [Online]. Available:
  \url{http://dx.doi.org/10.22331/q-2022-01-11-621}
\BIBentrySTDinterwordspacing

\bibitem{hein}
\BIBentryALTinterwordspacing
M.~Hein, J.~Eisert, and H.~J. Briegel, ``Multiparty entanglement in graph
  states,'' \emph{Phys. Rev. A}, vol.~69, p. 062311, Jun 2004. [Online].
  Available: \url{https://link.aps.org/doi/10.1103/PhysRevA.69.062311}
\BIBentrySTDinterwordspacing

\bibitem{PirRicOtt_17}
\BIBentryALTinterwordspacing
S.~Pirandola, R.~Laurenza, C.~Ottaviani, and L.~Banchi, ``Fundamental limits of
  repeaterless quantum communications,'' \emph{Nature Communications}, vol.~8,
  no.~1, Apr 2017. [Online]. Available:
  \url{http://dx.doi.org/10.1038/ncomms15043}
\BIBentrySTDinterwordspacing

\bibitem{BriegelDur}
\BIBentryALTinterwordspacing
H.-J. Briegel, W.~D\"ur, J.~I. Cirac, and P.~Zoller, ``Quantum repeaters: The
  role of imperfect local operations in quantum communication,'' \emph{Phys.
  Rev. Lett.}, vol.~81, pp. 5932--5935, Dec 1998. [Online]. Available:
  \url{https://link.aps.org/doi/10.1103/PhysRevLett.81.5932}
\BIBentrySTDinterwordspacing

\bibitem{Rodney}
R.~Van~Meter, \emph{Quantum Networking}, 1st~ed.\hskip 1em plus 0.5em minus
  0.4em\relax Wiley-IEEE Press, 2014.

\bibitem{Muralidharan2016OptimalAF}
S.~Muralidharan, L.~Li, J.~Kim, N.~L{\"u}tkenhaus, M.~D. Lukin, and L.~Jiang,
  ``Optimal architectures for long distance quantum communication,''
  \emph{Scientific Reports}, vol.~6, 2016.

\bibitem{duer-2019}
\BIBentryALTinterwordspacing
J.~Wallnöfer, A.~Pirker, M.~Zwerger, and W.~Dür, ``Multipartite state
  generation in quantum networks with optimal scaling,'' \emph{Scientific
  Reports}, vol.~9, no.~1, Jan 2019. [Online]. Available:
  \url{http://dx.doi.org/10.1038/s41598-018-36543-5}
\BIBentrySTDinterwordspacing

\bibitem{Dur-2019}
J.~Walln{\"o}fer, A.~Pirker, M.~Zwerger, and W.~D{\"u}r, ``Multipartite state
  generation in quantum networks with optimal scaling,'' \emph{Scientific
  Reports}, vol.~9, 2019.

\bibitem{interface}
\BIBentryALTinterwordspacing
H.~de~Riedmatten, M.~Afzelius, M.~U. Staudt, C.~Simon, and N.~Gisin, ``A
  solid-state light–matter interface at the single-photon level,''
  \emph{Nature}, vol. 456, no. 7223, p. 773–777, Dec 2008. [Online].
  Available: \url{http://dx.doi.org/10.1038/nature07607}
\BIBentrySTDinterwordspacing

\bibitem{interface2}
\BIBentryALTinterwordspacing
S.-J. Yang, X.-J. Wang, X.-H. Bao, and J.-W. Pan, ``An efficient quantum
  light–matter interface with sub-second lifetime,'' \emph{Nature Photonics},
  vol.~10, no.~6, p. 381–384, Apr 2016. [Online]. Available:
  \url{http://dx.doi.org/10.1038/nphoton.2016.51}
\BIBentrySTDinterwordspacing

\bibitem{Interface3}
\BIBentryALTinterwordspacing
L.~Bergeron, C.~Chartrand, A.~T.~K. Kurkjian, K.~J. Morse, H.~Riemann, N.~V.
  Abrosimov, P.~Becker, H.-J. Pohl, M.~L.~W. Thewalt, and S.~Simmons,
  ``Silicon-integrated telecommunications photon-spin interface,'' \emph{PRX
  Quantum}, vol.~1, p. 020301, Oct 2020. [Online]. Available:
  \url{https://link.aps.org/doi/10.1103/PRXQuantum.1.020301}
\BIBentrySTDinterwordspacing

\bibitem{walter-2017}
M.~Walter, D.~Gross, and J.~Eisert, ``Multi-partite entanglement,'' 2017.

\bibitem{masoud2021}
\BIBentryALTinterwordspacing
M.~Gharahi and S.~Mancini, ``Algebraic-geometric characterization of tripartite
  entanglement,'' \emph{Physical Review A}, vol. 104, no.~4, Oct 2021.
  [Online]. Available: \url{http://dx.doi.org/10.1103/PhysRevA.104.042402}
\BIBentrySTDinterwordspacing

\bibitem{masoud2020}
\BIBentryALTinterwordspacing
M.~Gharahi, S.~Mancini, and G.~Ottaviani, ``Fine-structure classification of
  multiqubit entanglement by algebraic geometry,'' \emph{Physical Review
  Research}, vol.~2, no.~4, Oct 2020. [Online]. Available:
  \url{http://dx.doi.org/10.1103/PhysRevResearch.2.043003}
\BIBentrySTDinterwordspacing

\bibitem{GME-Huber}
\BIBentryALTinterwordspacing
Z.-H. Ma, Z.-H. Chen, J.-L. Chen, C.~Spengler, A.~Gabriel, and M.~Huber,
  ``Measure of genuine multipartite entanglement with computable lower
  bounds,'' \emph{Physical Review A}, vol.~83, no.~6, Jun 2011. [Online].
  Available: \url{http://dx.doi.org/10.1103/PhysRevA.83.062325}
\BIBentrySTDinterwordspacing

\bibitem{ENtanglement-Siewert}
\BIBentryALTinterwordspacing
C.~Eltschka and J.~Siewert, ``Quantifying entanglement resources,''
  \emph{Journal of Physics A: Mathematical and Theoretical}, vol.~47, no.~42,
  p. 424005, Oct 2014. [Online]. Available:
  \url{http://dx.doi.org/10.1088/1751-8113/47/42/424005}
\BIBentrySTDinterwordspacing

\bibitem{li-2017}
M.~Li, J.~Wang, S.~Shen, Z.~Chen, and S.-M. Fei, ``Detection and measure of
  genuine tripartite entanglement with partial transposition and realignment of
  density matrices,'' \emph{Scientific reports}, vol.~7, no.~1, pp. 1--6, 2017.

\end{thebibliography}

\appendices

\section{Entanglement measures}
\label{appendix.1}

Quantum correlations has always been considered as a resource to perform tasks that are unachievable through classical resources, or to enhance other ones. Hence, a characterization and quantification of quantum correlations, in particular entanglement, has been widely studied \cite{Horodecki_2009,walter-2017}.\\
For a two-qubit state with density matrix $\rho$, quantum entanglement is completely characterized by \textit{concurrence}\cite{nielsen_chuang_2010,Horodecki_2009} $C(\rho))=max\{0,\mu_1-\mu-\mu_3-\mu_4\}$, where $\{ \mu_i\}$ denotes the set eigenvalues -- in decreasing order -- of the operator $\rho \tilde{\rho}$, with $\tilde{\rho}=(\sigma_y\otimes\sigma_y)~\rho~(\sigma_y\otimes\sigma_y)$. Concurrence $C(\cdot)$ is an entanglement monotone metric, with value equal to $1$ for maximally entangled states and value equal to $0$ for separable states.\\
For tripartite systems, entanglement measures are more intricate, and are only analytically found for special classes of states \cite{walter-2017, masoud2021,masoud2020}. For genuinely multipartite entangled (GME) states -- i.e, states that are not separable for any bipartition -- a known entanglement measure is the \textit{GME concurrence}\cite{GME-Huber,ENtanglement-Siewert,li-2017} $C_{\text{GME}}(\rho) =  \sqrt{2 \min \{ 1-\mathrm{Tr}(\rho_1^2),1-\mathrm{Tr}(\rho_2^2),1-\mathrm{Tr}(\rho_3^2) \} }$, with $\rho_i \eqdef Tr_{jk}(\rho)$ (with $j,k\neq i$) denoting the reduced density matrix for the $i$-th subsystem.

\section{Bell states generation}
\label{appendix.2}
\subsection*{Conditions for entanglement}
\begin{theorem}
\label{Theo:01}
The states in \eqref{Eq:04} are maximally entangled bipartite states if and only if the conditions in \eqref{Eq:05} hold.
\begin{proof}
The proof follows by reasoning as in Theorem~\ref{Theo:02}.
\end{proof}
\end{theorem}

\begin{proposition}
\label{Prop:01}
The states in \eqref{Eq:04} are bi-separable if and only if the conditions: $\exists \, i \in \{ 0,1\} \; : \; \overleftarrow{U_i} \ket{\varphi_i} = \overrightarrow{U_i} \ket{\varphi_i}$ hold.
\begin{proof}
The proof follows by reasoning as in Proposition~\ref{Prop:02}.
\end{proof}
\end{proposition}

\subsection*{Concurrence of the generated stated}
Here we detail the example provided to illustrate the results discussed in \ref{Sec:2.1}. 

We consider the unitaries given as $U=U'=\sigma_3$ and $\tilde{U}=\tilde{U'}=\exp{-i\sigma_2\lambda}$. In addition, we consider the initial product state $\ket{\psi}\ket{\phi}=\ket{\eta}\ket{\eta}$  where 
 \begin{equation*}
     \ket{\eta}=\sqrt{\alpha}\ket{0}+\sqrt{1-\alpha}\ket{1} 
     \label{initial}
 \end{equation*}
 The conditions for producing separable  states in Proposition~\ref{Prop:01}. are explicitly translated to
 \begin{align}
      &\bra{\eta}\sigma_3\exp(+i\sigma_2\lambda)\sigma_3\exp(-i\sigma_2\lambda)\ket{\eta}=\pm 1\nonumber\\
     &\implies \cos^2{\lambda}-\sin^2{\lambda}=\pm1\nonumber\\
     &\implies \lambda=\frac{\pi}{2}\hspace{4pt}or\hspace{4pt}\lambda=0\hspace{4pt} for\hspace{4pt}\lambda\in[0,\frac{\pi}{2}]
 \end{align}
 Only at these values of the unitary's parameter $\lambda$ both states generated by the quantum switch are separable according to Proposition.~\ref{Prop:01}.
Similarly, the conditions for maximal entangled states given in Theorem~\ref{Theo:01} are translated as
\begin{align}
&\bra{\eta}\sigma_3\exp(-i\sigma_2\lambda)\sigma_3\exp(-i\sigma_2\lambda)\ket{\eta}=0\nonumber\\
&\implies \cos^2{\lambda}-\sin^2{\lambda}=0\nonumber\\
&\implies \lambda=\frac{\pi}{4}\hspace{4pt} for\hspace{4pt}\lambda\in[0,\frac{\pi}{2}]
\label{condition-epr}
\end{align}
Only when $\lambda=\frac{\pi}{4}$ that both states generated by the switch are maximally entangled. The previous conditions surprisingly are 
independent of the parameter $\alpha$ of the state $\ket{\eta}$ and only depend on the parameter $\lambda$ of the unitary $\tilde{U}$.
In order to further check if these conditions hold, we study the entanglement content of the emerging states \eqref{Eq:04} for the chosen unitaries $U$ and $\tilde{U}$ and the product state $\ket{\eta}\ket{\eta}$. These states are given explicitly by
\begin{align*}
\psi_{+}^{(2)}&=\frac{1}{3+\cos(4\lambda)}\Big[[(2 \alpha -1) \cos (2 \lambda )
+1]\ket{00}\nonumber\\
&+[-2 \sqrt{(\alpha -1) \alpha } \cos (2 \lambda )]\ket{01}\nonumber\\
&+[-2 \sqrt{(\alpha -1) \alpha } \cos (2 \lambda )]\ket{10}\nonumber\\
&+[(1-2 \alpha ) \cos (2 \lambda )+1]\ket{11}\Big]\nonumber\\
\psi_{-}^{(2)}&=\frac{1}{\sqrt{2\sin^2{2\lambda}}}\Big[4\cos{\lambda}\sin{\lambda}\sqrt{(1-\alpha)\alpha}\big(\ket{00}-\ket{11}\big)\nonumber\\
&+(2\alpha-1)\sin{2\lambda}\big(\ket{01}+\ket{10}\big)\Big]
\end{align*}

The concurrence of the previous states  is given respectively by 
\begin{align*}
	C(\psi_+^{(2)})&=\frac{1}{2}\Bigg[\frac{3-4\cos(4\lambda)+\cos(8\lambda)}{(3+\cos(4\lambda))^2}\Bigg] \label{concurrence+}\\
	C(\psi_-^{(2)})&=
    \begin{cases}
      1 & \text{if $\lambda\in(0,\frac{\pi}{2})$}\\
      0 & \text{if $\lambda=0$ or $\lambda=\frac{\pi}{2}$}
    \end{cases}
\end{align*}

\section{GHZ-like states}
\subsection*{Conditions for entanglement}
\label{appendix.3}
Here we derive in Theorem~\ref{Theo:02} the necessary and sufficient condition for generating a 3-partite GHZ-like state through superposition of causal orders. 
\begin{theorem}
\label{Theo:02}
The states in \eqref{Eq:06} are GHZ-like states if and only if the condition given in \eqref{Eq:07} holds.
\begin{proof}
We first observe that any tripartite GHZ-like state $\ket{\Psi^{(3)}}$ is equivalent to the GHZ state $\ket{\text{GHZ}} = \frac{1}{\sqrt{2}}(\ket{000}+\ket{111})$ through local unitaries \cite{walter-2017}  -- i.e.,  $\ket{\text{GHZ}} = (U_0^{\text{LU}} \otimes U_1^{\text{LU}} \otimes U_2^{\text{LU}}) \ket{\Psi^{(3)}}$.\\
Case $\Longleftarrow$ (sufficient condition). By hypothesis, $\overleftarrow{U_i} \ket{\varphi_i} = \left( \overrightarrow{U_i} \ket{\varphi_i} \right)^\perp$ for any $i$. Hence, \eqref{Eq:06} is equivalent to:

\begin{align}
\ket{\psi^{(3)}_\pm} &= \frac{1}{\sqrt{L}_\pm} \overrightarrow{U_0} \ket{\varphi_0} \overrightarrow{U_1} \ket{\varphi_1} \overrightarrow{U_2} \ket{\varphi_2} \nonumber\\
&\pm (\overrightarrow{U_0} \ket{\varphi_0})^\perp ( \overrightarrow{U_1}\ket{\varphi_1})^\perp ( \overrightarrow{U_2}\ket{\varphi_2} )^\perp
\end{align}
By defining local unitaries such that $U_i^{\text{LU}} \overrightarrow{U_i} \ket{\varphi_i} = \ket{0}$ (hence, $U_i^{\text{LU}} \left( \overrightarrow{U_i} \ket{\varphi_i} \right)^\perp = \ket{1}$), the thesis follows.\\

Case $\Longrightarrow$ (necessary condition). By hypothesis, \eqref{Eq:06} is a GHZ-like state. We prove the case with a \textit{reductio ad absurdum} by supposing that there exists at least one $i$, say $i=0$, so that $\overleftarrow{U_0} \ket{\varphi_0} \neq \left( \overrightarrow{U_0} \ket{\varphi_0} \right)^\perp$. From GHZ-like state definition, there exist $U_1^{\text{LU}},U_2^{\text{LU}}$ such that $\ket{\psi^{(3)}_\pm}$ is LU-equivalent to the following (by neglecting the normalization factor for the sake of simplicity):
\begin{equation}
    \label{eq:C.2}
    \left( I \otimes U_1^{\text{LU}} \otimes U_2^{\text{LU}} \right) \ket{\psi^{(3)}_\pm} = \overrightarrow{U_0} \ket{\varphi_0} \otimes  \ket{00} \pm \overleftarrow{U_0} \ket{\varphi_0} \otimes \ket{11}
\end{equation}
Then, there must exist another unitary $U_0^{\text{LU}}$ acting on the first qubit such that \eqref{eq:C.2} is equivalent to the $\ket{GHZ}$ state, i.e.:
\begin{align}
    \label{eq:C.3}
    &\left( U_0^{\text{LU}} \otimes I \otimes I \right) \left(\overrightarrow{U_0} \ket{\varphi_0} \otimes \ket{00} \pm \overleftarrow{U_0} \ket{\varphi_0} \otimes \ket{11}\right) \nonumber\\
    &= \frac{1}{\sqrt{2}}(\ket{000} + \ket{111})
\end{align}
Since unitary matrices preserve orthogonality, from \eqref{eq:C.3}, it follows that $\overrightarrow{U_0}\ket{\varphi_0} = \left( \overleftarrow{U_0} \ket{\varphi_0} \right)^\perp$, which constitutes a \textit{reductio ad absurdum}. Hence, the thesis follow.
\end{proof}
\end{theorem}
\begin{remark}
Clearly, the above result can be straightforwardly extended to $n$-partite GHZ-like states $\ket{\psi^{(n)}_\pm}$ generated through an even superposition of the two alternative causal orders between two $n$-qubit local unitaries $V^{(n)} = \bigotimes_{i=0}^{n-1} U_i$ and $\tilde{V}^{(n)}= \bigotimes_{i=0}^{n-1} \tilde{U_i}$ acting on an initially pure product $n$-partite state $\bigotimes_{i=0}^{n-1} \ket{\varphi_i}$
by following the same reasoning and, in such a case, the necessary and sufficient condition becomes: $\bra{\varphi_i} \overleftarrow{U_i}^{\dagger} \overrightarrow{U_i} \ket{\varphi_i} = 0 \; \forall \, i=0, \ldots, n-1$.
\end{remark}
\begin{proposition}
\label{Prop:02}
The states in \eqref{Eq:06} are bi-separable if and only if the condition given in \eqref{Eq:08} holds.
\begin{proof}
In the following, we directly prove the proposition for the arbitrary states $\ket{\psi^{(n)}_\pm}$ generated through an even superposition of the two alternative causal orders between two $n$-qubit local unitaries $V^{(n)} = \bigotimes_{i=0}^{n-1} U_i$ and $\tilde{V}^{(n)}= \bigotimes_{i=0}^{n-1} \tilde{U_i}$ acting on an initially pure product $n$-partite state $\bigotimes_{i=0}^{n-1} \ket{\varphi_i}$.\\
Case $\Longleftarrow$ (sufficient condition). It is straightforward to recognize that, whenever there exists at least an $i$ such that $\overleftarrow{U_i} \ket{\psi_i} = \overrightarrow{U_i} \ket{\psi_i}$, the states $\ket{\psi^{(n)}_\pm}$ in Eq.~\ref{Eq:06} are separable.\\
Case $\Longrightarrow$ (necessary condition). By hypothesis $\ket{\psi^{(n)}_\pm}$ are separable. Hence, there exist a partition so that $\ket{\psi^{(n)}_\pm} = \ket{\psi_A} \otimes \ket{\psi_B}$ with $\ket{\psi_A}$ being pure state of the first subsystem $A$. Let us assume, without loss of generality, subsystem $A$ consisting of the first two qubits of \eqref{Eq:06}. We prove the case with a \textit{reductio ad absurdum} by supposing that $\overleftarrow{U_i} \ket{\psi_i} \neq \overrightarrow{U_i} \ket{\psi_i}$ for any $i=\{0,1,\ldots,n-1\}$. Since $\ket{\psi^{(n)}_\pm}$ are separable, there exist two local unitaries acting on the first two qubits\footnote{The same reasoning -- as well as the same result -- holds by considering local unitaries acting on the remaining $n-2$ qubits.} such that:

\begin{strip}
\rule[-1ex]{\columnwidth}{1pt}\rule[-1ex]{1pt}{1.5ex}
    \label{eq:C.4}
\begin{align}    
    & \left( U_0^{\text{LU}} \otimes U_1^{\text{LU}} \otimes \underbrace{I \ldots \otimes I}_{n-2}  \right)\ket{\psi^{(n)}_\pm}
        = \nonumber \\
        & \quad = U_0^{\text{LU}} \overrightarrow{U_0} \ket{\varphi_0} \otimes U_1^{\text{LU}} \overrightarrow{U_1} \ket{\varphi_1} \otimes \left(\overrightarrow{U_3} \otimes \ldots \otimes \overrightarrow{U}_{n-1}\right)\ket{\varphi_3 \ldots \varphi_{n-1}} \pm \nonumber \\
        & \quad \quad \quad U_0^{\text{LU}} \overleftarrow{U_0} \ket{\varphi_0} \otimes U_1^{\text{LU}} \overleftarrow{U_1} \ket{\varphi_1} \otimes \left(\overleftarrow{U_3} \otimes \ldots \otimes \overleftarrow{U}_{n-1}\right)\ket{\varphi_3 \ldots \varphi_{n-1}} \nonumber \\
        & \quad = \ket{\uparrow \uparrow} \otimes \left(\overrightarrow{U_3} \otimes \ldots \otimes \overrightarrow{U}_{n-1}\right)\ket{\varphi_3 \ldots \varphi_{n-1}} \pm c \ket{\uparrow \uparrow} \otimes \left(\overleftarrow{U_3} \otimes \ldots \otimes \overleftarrow{U}_{n-1} \right) \ket{\varphi_3 \ldots \varphi_{n-1}}
\end{align}
\hfill\rule[1ex]{1pt}{1.5ex}\rule[2.3ex]{\columnwidth}{1pt}
\end{strip}

with $\ket{\uparrow \uparrow}$ denoting a certain state for subsystem $A$. Since unitary matrices preserve inner product, from \eqref{eq:C.4} it results that $\overleftarrow{U_0} \otimes \overleftarrow{U_1} \ket{\varphi_0 \varphi_1} = c \, \overrightarrow{U_0} \otimes \overrightarrow{U_1} \ket{\varphi_0 \varphi_1}$. But this last equality requires that $\overleftarrow{U_i} \ket{\psi_i} = \overrightarrow{U_i} \ket{\psi_i}$ for any $i=\{0,1\}$, which constitutes a \textit{reductio ad absurdum}. Hence, the thesis follow.
\end{proof}
\end{proposition}

\subsection*{The GME concurrence of the generated states}
we consider the unitaries used previously -- $U=\sigma_3$ and $\tilde{U}=\exp{-i\lambda \sigma_2}$ -- as well as the initial product state $\ket{\eta}\ket{\eta}\ket{\eta}$ where $\ket{\eta}=\sqrt{\alpha}\ket{0}+\sqrt{1-\alpha}\ket{1}$. Quantum switching the local unitaries $U\otimes U\otimes U$ and $\tilde{U}\otimes \tilde{U}\otimes \tilde{U}$ and letting them act on $\ket{\eta}\ket{\eta}\ket{\eta}$ we get, after measuring the control qubit on the coherent basis, the emerging states  
\begin{align}
    \ket{\psi_+^{(3)}}&=\frac{1}{L}\Big(a\ket{000}+b(\ket{001}+\ket{010}+\ket{100})\nonumber\\
    &+e(\ket{110}+\ket{101}+\ket{011}+h\ket{111})\Big)\nonumber\\
    \ket{\psi_-^{(3)}}&=\frac{1}{L'}\Big(a'\ket{000}+b'(\ket{001}+\ket{010}+\ket{100})\nonumber\\
    &+e'(\ket{110}+\ket{101}+\ket{011}+h'\ket{111})\Big)
\label{emerging-tripartite}
\end{align}
where the parameters $a, b, e, h, L$ and $a', b', e', h', L'$ are given by
\begin{align}
	&a=-\sqrt{\alpha } \cos (\lambda ) ((4 \alpha -3) \cos (2 \lambda )-2 \alpha +3)\nonumber\\
	& b=\sqrt{1-\alpha } \cos (\lambda ) ((4 \alpha -1) \cos (2 \lambda )-2 \alpha +1)\nonumber\\
	& e=\sqrt{\alpha } \cos (\lambda ) ((4 \alpha -3) \cos (2 \lambda )-2 \alpha +1)\nonumber\\
	& h= -\sqrt{1-\alpha }\cos{\lambda} ((4 \alpha -1) \cos (2 \lambda )-1-2\alpha)\nonumber\\
	&L=\sqrt{\frac{1}{2} (3 \cos (2 \lambda )+\cos (6 \lambda )+4)}\nonumber\\
 \end{align}
 \begin{align}
	&a'=-\sqrt{1-\alpha } \sin (\lambda ) ((4 \alpha -1) \cos (2 \lambda )+1+2\alpha)\nonumber\\
	& b'=\sqrt{\alpha } \sin (\lambda ) ((4 \alpha -3) \cos (2 \lambda )+2 \alpha -3)\nonumber\\
	& e'=-\sqrt{\alpha }\sin{\lambda} ((4 \alpha -3) \cos (2 \lambda )-1+2\alpha)\nonumber\\
	& h'= \sqrt{1-\alpha }\sin{\lambda} ((4 \alpha -1) \cos (2 \lambda )-1+2\alpha)\nonumber\\
	&L'=\sqrt{\frac{1}{2} (4-3 \cos (2 \lambda )-\cos (6 \lambda ))}
\end{align}

In order to study the biseparability of the state \eqref{emerging-tripartite}, we need to study the partial trace with respect to different bipartitions $\rho_{1\mid23}$, $\rho_{2\mid13}$, $\rho_{3\mid12}$ which turn out to be equal for the states $\ket{\psi_\pm^{(3)}}$. The reduced density matrices of  these latter are given by \ref{reduced1} and \ref{reduced2}.

\begin{figure*}[t]
\label{reduced-states}
\normalsize
\begin{align}
&\rho_+^{reduced}=\frac{1}{L^2}\left(
	\begin{array}{cc}
	\cos ^2(\lambda ) (4 (\alpha -1) \cos (2 \lambda )+\cos (4 \lambda )+3) & -4 \sqrt{-(\alpha -1) \alpha } \cos ^2(\lambda ) \cos (2 \lambda ) \\
	-4 \sqrt{-(\alpha -1) \alpha } \cos ^2(\lambda ) \cos (2 \lambda ) & \cos ^2(\lambda ) (-4 \alpha  \cos (2 \lambda )+\cos (4 \lambda )+3) \\
	\end{array}
	\right) \label{reduced1}\\
&\rho_-^{reduced}=\frac{1}{L'^2}\left(
	\begin{array}{cc}
	\sin ^2(\lambda ) (4 \alpha  \cos (2 \lambda )+\cos (4 \lambda )+3) & -4 \sqrt{(\alpha -1) \alpha } \sin ^2(\lambda ) \cos (2 \lambda ) \\
	-4 \sqrt{(\alpha -1) \alpha } \sin ^2(\lambda ) \cos (2 \lambda ) & \sin ^2(\lambda ) (-4 (\alpha-1)  \cos (2 \lambda )+\cos (4 \lambda )+3)  \\
	\end{array}
	\right)\label{reduced2}
\end{align}
\hrulefill
\end{figure*}

In turn, the GME concurrence for both states is written as 
\begin{align}
	&C_{GME}(\rho_+)=\frac{8 \sin ^4(\lambda ) (\cos (4 \lambda )+3)}{(-2 \cos (2 \lambda )+\cos (4 \lambda )+3)^2}\nonumber\\
	&C_{GME}(\rho_-)=\frac{8 \cos ^4(\lambda ) (\cos (4 \lambda )+3)}{(2 \cos (2 \lambda )+\cos (4 \lambda )+3)^2}
	\label{eq: tripartite GME}
\end{align}
\section{W-like states}
\label{appendix.4}
\subsection*{Conditions for entanglement}
\begin{theorem}
\label{Theo:03}
The states in \eqref{Eq:11} are W-like states if and only if the condition given in \eqref{Eq:12} holds.
\begin{proof}
We first note that any tripartite W-like state $\ket{\Psi^{(3)}}$ is equivalent to $\ket{W}=\frac{1}{\sqrt{3}}(\ket{100}+\ket{010}+\ket{001})$ by local unitaries -i.e., $\ket{W}=\big(U_0^{LU}\otimes U_1^{LU}\otimes U_2^{LU}\big)\ket{\Psi^{(3)}}$.
Similarly to the proof of Theorem.~2, the proof of the sufficiency of the condition in \eqref{Eq:12} is straightforward. In the meanwhile, the necessity can be proved by \textit{reductio ad absurdum} as in Theorem.~2, by supposing that there exists at least one $i$ -- say, without loss of generality, $i=0$ -- so that $\overleftarrow{U}_0\ket{\varphi}\neq \big(\overrightarrow{U}_0\ket{\varphi_0}\big)^{\perp}$. From the W-like state equivalence, there exists a local unitary of the form $I\otimes U_1^{LU}\otimes U_2^{LU}$ such that 
\begin{align}
    \label{eq:23}
    &\left( I \otimes U_1^{\text{LU}} \otimes U_2^{\text{LU}} \right) \ket{\psi^{(3)}_{\pm\pm}} =\nonumber\\
    &\overleftarrow{U_0} \ket{\varphi_0} \otimes  \ket{00} \pm \overrightarrow{U_0} \ket{\varphi_0} \otimes \ket{10} \pm \overrightarrow{U_0} \ket{\varphi_0} \otimes \ket{01}
\end{align}
Then, there must exist another unitary $U_0^{\text{LU}}$ acting on the first qubit such that \eqref{eq:23} is equivalent to W state, i.e.:
\begin{strip}
\rule[-1ex]{\columnwidth}{1pt}\rule[-1ex]{1pt}{1.5ex}
\begin{align}
    \label{eq:24}
    &\left( U_0^{\text{LU}} \otimes I \otimes I \right) \overleftarrow{U_0} \ket{\varphi_0} \otimes  \ket{00} \pm \overrightarrow{U_0} \ket{\varphi_0} \otimes \ket{10} \pm \overrightarrow{U_0} \ket{\varphi_0} \otimes \ket{01} = \frac{1}{\sqrt{3}}\big(  \ket{100} \pm \ket{010} \pm \ket{001}\big)
\end{align}
\hfill\rule[1ex]{1pt}{1.5ex}\rule[2.3ex]{\columnwidth}{1pt}
\end{strip}
Since unitary matrices preserve orthogonality, from \eqref{eq:24}, it follows that $\overrightarrow{U_0}\ket{\varphi_0} = \left( \overleftarrow{U_0} \ket{\varphi_0} \right)^\perp$, which constitutes a \textit{reductio ad absurdum}. Hence, the thesis follow.
\end{proof}
\end{theorem}

\begin{remark}
Although the deterministic generation of any $n$-partite GHZ-like state requires only a qubit degree of freedom controlling two different evolutions coherently, the deterministic generation of W-like states requires a higher-order control of the causal orders. Specifically, $n$ local unitaries --  with $n$ being a power of $2$ -- must be arranged in a particular way. The rationale for this requirement lays in the necessity of having a maximally coherent basis of states that serves as a measurement setup on the controlling degrees of freedom, allowing the deterministic generation of the superposition required in the W states on all outputs. This requirement can only be met in Hilbert spaces of dimension which is a power of two. In such space, an orthonormal basis of maximally coherent states exists, and it can be used to coherentely control the order of the local unitaries. Therefore, generating $n$-partite W-like states deterministically, where $n=2^d$, encounters no problem as we can always find a maximally coherent orthonormal basis achieving this. Instead, a slight adjustment on the control strategy needs to be handled in order to achieve the deterministic generation of $n$-partite W states when $n$ is not a power of $2$. To overcome this issue, we embed the control degrees of freedom in a larger Hilbert space of dimension $2^d$ where $d=\lceil \log_2 n \rceil$. \textcolor{black}{It is important to note that this requirement is not necessary if only heralded entanglement generation is desired. In this case, any qudit state of dimension $n$ can be used to control the order of the $n$ local unitaries. Otherwise, this condition is not suitable for the deterministic generation of entangled states.}
\end{remark}

\begin{proposition}
\label{Prop:03}
The state in \eqref{Eq:11} is bi-separable if and only if the condition given in \eqref{Eq:13} holds.
\begin{proof}
The proof follows the similar steps of Proposition~\ref{Prop:02}.
\end{proof}
\end{proposition}

\subsection*{The deterministic generation of 3-W states}
In this section we show how the states in (11) are obtained from a superposition of causal order. 
By considering the Kraus operator of the model to be given by the unitary   
\begin{align}
    S =& \big(\tilde{U_0} \otimes U_1 \otimes U_2 \cdot U_0 \otimes \tilde{U_1} \otimes \tilde{U_2}\big)\otimes \ket{1}\bra{1}_c\nonumber\\
    &+\left( U_0 \otimes \tilde{U_1} \otimes U_2 \cdot \tilde{U_0} \otimes U_1 \otimes \tilde{U_2} \right)\otimes\ket{2}\bra{2}_c\nonumber\\
    &+\left( U_0 \otimes U_1 \otimes \tilde{U_2} \cdot \tilde{U_0} \otimes \tilde{U_1} \otimes U_2 \right) \otimes \ket{3}\bra{3}_c
\end{align}    
This acts on a joint system-control state  $\big(\ket{\varphi_0 \varphi_1 \varphi_2}\big)\otimes \ket{+^{(3)}}_c$ with 
\begin{equation}
    \ket{+^{(3)}}_c=\frac{1}{\sqrt{3}}\big(\ket{0}_c+\ket{1}_c+\ket{2}_c\big)
\end{equation}
The resulting joint evolution is given by
\begin{align*}
   \frac{1}{\sqrt{3}} [ & ( \tilde{U_0} \otimes U_1 \otimes U_2 \cdot U_0 \otimes \tilde{U_1} \otimes \tilde{U_2} ) \ket{\varphi_0 \varphi_1 \varphi_2} \otimes \ket{0}_c \nonumber\\
   & +( U_0 \otimes \tilde{U_1} \otimes U_2 \cdot \tilde{U_0} \otimes U_1 \otimes \tilde{U_2} ) \ket{\varphi_0 \varphi_1 \varphi_1} \otimes \ket{1}_c   \nonumber \\
    & +( U_0 \otimes U_1 \otimes \tilde{U_2} \cdot \tilde{U_0} \otimes \tilde{U_1} \otimes U_2) \ket{\varphi_1 \varphi_1 \varphi_2} \otimes \ket{3}_c ] 
\end{align*}

Because of the issue of the measurement discussed in (Discussion) an encoding of the control in a higher dimensional Hilbert space whose dimension is a power of $2$ is needed, in order to find a maximally coherent basis.  
If we perform the encoding 
\begin{align*}
    \ket{0}\rightarrow \ket{00}\\
    \ket{1}\rightarrow \ket{01}\\
    \ket{2}\rightarrow \ket{10}\\
\end{align*}
it allows us to perform a measurement in the maximally coherent basis given as
\begin{align*}
  \ket{++}= \frac{1}{\sqrt{4}}\big(\ket{00}+\ket{01}+\ket{10}+\ket{11}\big)\\ 
  \ket{+-}= \frac{1}{\sqrt{4}}\big(\ket{00}-\ket{01}+\ket{10}-\ket{11}\big)\\ 
  \ket{-+}= \frac{1}{\sqrt{4}}\big(\ket{00}+\ket{01}-\ket{10}-\ket{11}\big)\\
  \ket{--}= \frac{1}{\sqrt{4}}\big(\ket{00}-\ket{01}-\ket{10}+\ket{11}\big)\\ 
\end{align*}
This measurement procedure generates the following states
\begin{align*}
    \ket{\psi^{(3)}_{\pm\pm}} &=\frac{1}{\sqrt{3}} [ \left( \tilde{U_0} \otimes U_1 \otimes U_2 \cdot U_0 \otimes \tilde{U_1} \otimes \tilde{U_2} \right)  \ket{\varphi_0 \varphi_1 \varphi_2} \nonumber\\
    &\pm \left( U_0 \otimes \tilde{U_1} \otimes U_2 \cdot \tilde{U_0} \otimes U_1 \otimes \tilde{U_2} \right) \ket{\varphi_0 \varphi_1 \varphi_1}  \nonumber \\
    & \pm \left( U_0 \otimes U_1 \otimes \tilde{U_2} \cdot \tilde{U_0} \otimes \tilde{U_1} \otimes U_2 \right) \ket{\varphi_1 \varphi_1 \varphi_2}]  
\end{align*}


\section{Graph states}
\label{appendix.5}
Here we prove that the indefinite causal strategy $S$ given in \eqref{eq:22} does generate of a graph state $\ket{G}$, by assuming the knowledge of the corresponding graph $G=(V,E)$. From graph state definition, we have:
\begin{equation}
    \ket{G}=\otimes_{(i,j)\in E} CZ^{(i,j)}\ket{+}^{n} \label{eq:23bis}
\end{equation}
The state resulting from the strategy $S$ in \eqref{eq:22}, after measuring the control qubits in the coherent basis, is given by: 
\begin{equation}
    \ket{G}=\otimes_{(i,j)\in E} SCO_\pm^{(i,j)}\ket{\eta}^{n} \label{eq:24bis}
\end{equation}
with $\ket{\eta}$ denoting the input state and $SCO_\pm^{(i,j)}=V^{(i,j)}\tilde{V}^{(i,j)}\pm\tilde{V}^{(i,j)}V^{(i,j)}$. We note that it does not exist any local unitary operation mapping the two operations $CZ$ and $SCO$ each others, i.e, they are not equivalent up to a local unitary. If this would be true, we should be able to find a local two-qubit unitary $T\otimes W$ satisfying the equation: 
\begin{equation}
    (T\otimes W)CZ(T^{\dagger}\otimes W^{\dagger})=SCO  
\end{equation}
with $T$ and $W$ arbitrary single-qubit unitaries. Clearly, this equation has no solution since $\mathrm{Tr}((T\otimes W)CZ(T^{\dagger}\otimes W^{\dagger}))=\mathrm{Tr}(CZ)=2\neq \mathrm{Tr}(SCO_\pm)$. 

Luckily, the similarity between $CZ$ and $SCO_\pm$ can be retrieved, effictively, by their respective action on separable input states. By fixing the input states $\ket{\eta}$, we can find a post-processing local unitary that makes the two operations generating LU equivalent states. 
In fact, if we set the input state $\ket{\eta}=\ket{0}$, we obtain:
\begin{align}
    \label{eq:27}
    SCO^{+}\ket{00}&=\frac{1}{\sqrt{2}}\Big(\ket{00}+\ket{11}\Big)\nonumber\\
    SCO^{-}\ket{00}&=\frac{1}{\sqrt{2}}\Big(\ket{10}+\ket{01}\Big)
\end{align}
On the other hand, it results:
\begin{equation}
    \label{eq:28}
    CZ\ket{++}=\frac{1}{\sqrt{2}}\Big(\ket{0+}+\ket{1-}\Big)
\end{equation}
This shows that the states in \eqref{eq:23bis} and \eqref{eq:24bis} are equivalent up to a local unitary, given by $I\otimes H$ and $X\otimes H$ depending on the outcomes $\{+,-\}$ on the control qubit respectively. 

Finally, to asses the necessary nature of conditions in \eqref{Eq:17}, we follow the same reasoning  of \textit{reductio ad absurdum} used in Theorem.~\ref{Theo:02}. Let us suppose that there exists an indefinite causal order strategy that generates $\ket{G}$ with the appropriate entanglement rank $r_{AB}=r'$ on bipartition $\{A,B\}$ and that violates one of the conditions in \eqref{Eq:17} corresponding to the same bipartition $\{A,B\}$, and let this condition be without any loss of generality the following one:
\begin{align}
    \bra{\psi}_y|\psi\rangle_{x}&=\Bigg[\sum_{{j'}_{x}}^{2^{l}}\Big[\otimes_{i\in A}\bra{0}_iU_i^{({j'}_{x})\dagger}\Big]\Bigg] \Bigg[\sum_{{j}_{y}}^{2^{l}}\Big[\otimes_{j\in A}U_i^{({j}_{y})}\ket{0}_i\Big]\Bigg]\nonumber\\
    &=1
\end{align}
In this case, the Schmidt decomposition in the bipartition $\{A,B\}$  violates the corresponding entanglement rank of the state and is given by:
\begin{equation}
    \ket{G}=\sum_{h=1}^{2^{r'}}\ket{\phi}_h\ket{\psi}_h=
    \sum_{h\neq x,y}^{2^{r'}}\ket{\phi}_h\ket{\psi}_h+(\ket{\phi}_x+\ket{\phi}_{y})\ket{\psi}_x
\end{equation}
and therefore the schmidt rank $r_{AB}=r'-1$, which contradicts the assumption. Hence, the conditions in \eqref{Eq:17} must be fulfilled to be able to generate a graph state with appropriate entanglement ranks. 

\color{black}

\begin{IEEEbiography}
[{\includegraphics[width=1in,height=1.25in,clip,keepaspectratio]{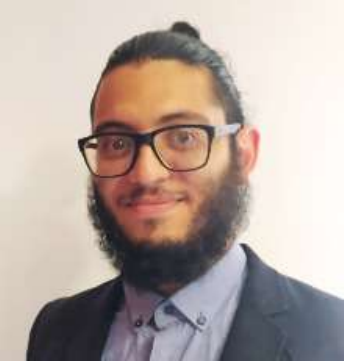}}]{Seid Koudia} Received the B.Sc degree in fundamental physics in 2015 and the M.Sc degree in theoretical physics with distinction in 2017 from the University of Sciences and Technology Houari Boumedien (USTHB). Currently, he is pursuing a PhD degree in Quantum technologies with the Future Communications Laboratory (FLY), Department of Electrical Engineering and Information Technology (DIETI). His research interests include quantum information theory, quantum communications, quantum networks and quantum coding theory.
\end{IEEEbiography}

\begin{IEEEbiography}
[{\includegraphics[width=1in,height=1.25in,clip,keepaspectratio]{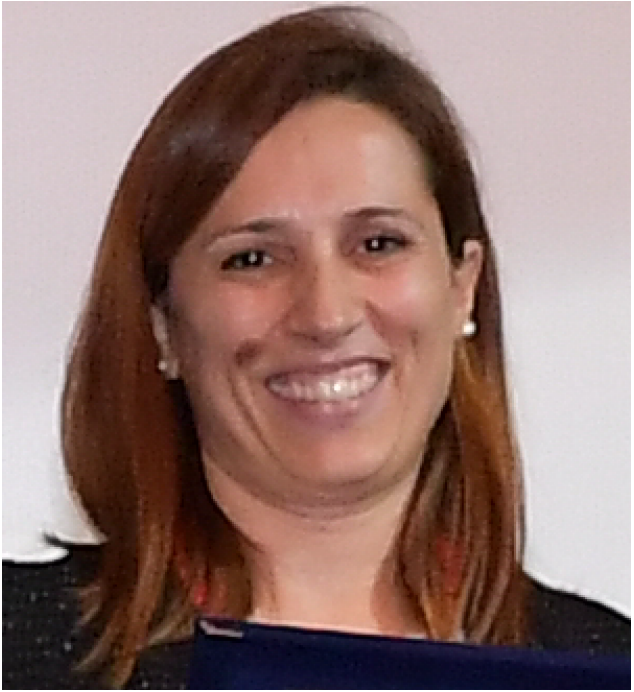}}]
{Angela Sara Cacciapuoti} (M'10, SM'16) is a professor at the University of Naples Federico II (Italy). Since July 2018 she held the national habilitation as “Full Professor” in Telecommunications Engineering. Her work has appeared in first tier IEEE journals and she has received different awards and recognition, including the \textit{``2022 IEEE ComSoc Best Tutorial Paper Award''} and \textit{``2021 N2Women: Stars in Networking and Communications''}. For the Quantum Internet topics, she is a \textit{IEEE ComSoc Distinguished Lecturer}, class of 2022-2023. Currently, Angela Sara serves as \textit{Area Editor} for IEEE Communications Letters, and as \textit{Editor/Associate Editor} for the journals: IEEE Trans. on Communications, IEEE Trans. on Wireless Communications, IEEE Trans. on Quantum Engineering, IEEE Network. She was the recipient of the \textit{2017 Exemplary Editor Award} of the IEEE Communications Letters. From 2020 to 2021, Angela Sara was the Vice-Chair of the IEEE ComSoc Women in Communications Engineering (WICE). Previously, she has been appointed as Publicity Chair of WICE. From 2016 to 2019 she has been an appointed member of the IEEE ComSoc Young Professionals Standing Committee. From 2017 to 2020, she has been the Treasurer of the IEEE Women in Engineering (WIE) Affinity Group of the IEEE Italy Section. Her current research interests are mainly in Quantum Communications, Quantum Networks and Quantum Information Processing.
\end{IEEEbiography}

\begin{IEEEbiography}
[{\includegraphics[width=1in,height=1.25in,clip,keepaspectratio]{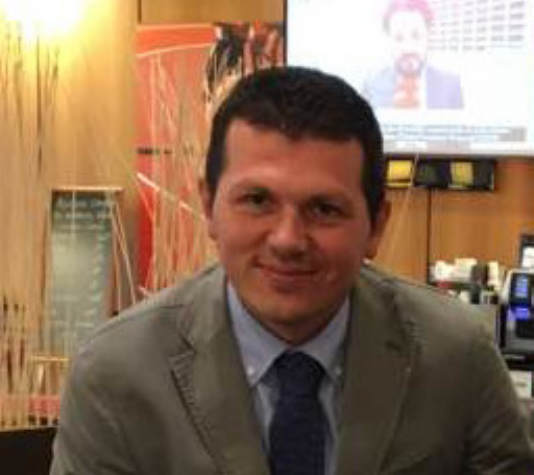}}]{Marcello Caleffi} (M'12, SM'16) received the M.S. degree with the highest score (summa cum laude) in computer science engineering from the University of Lecce, Lecce, Italy, in 2005, and the Ph.D. degree in electronic and telecommunications engineering from the University of Naples Federico II, Naples, Italy, in 2009.  Currently, he is Associate professor at the DIETI Department, University of Naples Federico II. From 2010 to 2011, he was with the Broadband Wireless Networking Laboratory at Georgia Institute of Technology, Atlanta, as visiting researcher. In 2011, he was also with the NaNoNetworking Center in Catalunya (N3Cat) at the Universitat Politecnica de Catalunya (UPC), Barcelona, as visiting researcher. Since July 2018, he held the Italian national habilitation as \textit{Full Professor} in Telecommunications Engineering. His work appeared in several premier IEEE Transactions and Journals, and he received multiple awards, including \textit{best strategy} award, \textit{most downloaded article} awards and \textit{most cited article} awards. Currently, he serves as \textit{associate technical editor} for IEEE Communications Magazine and as \textit{associate editor} for IEEE Trans. on Quantum Engineering and IEEE Communications Letters. He served as Chair, TPC Chair, Session Chair, and TPC Member for several premier IEEE conferences. In 2016, he was elevated to IEEE Senior Member and in 2017 he has been appointed as Distinguished Lecturer from the \textit{IEEE Computer Society}. In December 2017, he has been elected Treasurer of the Joint \textit{IEEE VT/ComSoc Chapter Italy Section}. In December 2018, he has been appointed member of the IEEE \textit{New Initiatives Committee}.
\end{IEEEbiography}
\EOD
\end{document}